\documentclass[12pt]{article}
\usepackage{epsfig}
\begin{document}
\title{Photon-induced production of the exotic charged charmonium-like state $Z_c(3900)$ off nuclear targets and its internal structure}
\author{E. Ya. Paryev\\
{\it Institute for Nuclear Research of the Russian Academy of Sciences}\\
{\it Moscow, Russia}}

\renewcommand{\today}{}
\maketitle

\begin{abstract}
In this paper, we investigate the possibility to study the famous charged charmonium-like state $Z_c(3900)$
production off nuclear targets and its properties in inclusive photon-induced reactions near the kinematic threshold within the collision model based on the nuclear spectral function. The model accounts for its charged components $Z_c(3900)^{\pm}$ production in direct photon--nucleon interactions as well as three different popular scenarios for their internal structures: compact tetraquarks, molecules of the two open-charm mesons and mixtures of both of them. We calculate the absolute and relative excitation functions for production of $Z_c(3900)^{\pm}$ mesons on $^{12}$C and $^{184}$W target nuclei at initial photon energies of 9.0--17.5 GeV, the absolute momentum differential cross sections and their ratios for the $Z_c(3900)^{\pm}$ production off these target nuclei at laboratory polar angles of 0$^{\circ}$--10$^{\circ}$ and at photon energy of 14 GeV as well as the A-dependences of the ratios of the total cross sections for $Z_c(3900)^{\pm}$ production at photon energy of 14 GeV within the adopted scenarios for the $Z_c(3900)^{\pm}$ internal structures. Our results in particular indicate that the total cross sections for $Z_c(3900)^{+}$$(Z_c(3900)^-)$ production have at above threshold photon energies of $\sim$ 13--17 GeV a well measurable strengths $\sim$ 50--200 (50--200) and 200--1500 (300--2000) nb for carbon and tungsten target nuclei, respectively. Therefore, one might expect to measure these strengths at the upgraded up to 22 GeV CEBAF facility. We also demonstrate that the absolute and relative observables considered show a certain sensitivity to the
$Z_c(3900)^{\pm}$ internal structures which are by far the best known. Hence, they might be useful for the determination of these structures -- the issue which has attracted much attention in the hadron physics community -- from the comparison of them with the experimental data from the future high-precision experiments at the above facility.
\end{abstract}

\newpage

\section*{1. Introduction}

\hspace{1.5cm} In the last two decades, high-energy experiments have reported a plethora of exotic narrow hadronic structures, which cannot be accommodated within the conventional quark model for quark-antiquark mesons and three-quarks baryons and which are composed of four or five quarks (and antiquarks) and are named as tetraquark and pentaquark states, respectively. Thus, many unconventional charmonium- and bottomonium-like states with hidden charm or beauty, charged or neutral, usually referred to as $XYZ$ states as well as the hidden-charm non-strange $P_c$ and strange $P_{cs}$ pentaquark states, the doubly-charmed tetraquark $T^+_{cc}(3875)$ state, the fully charmed tetraquark $X(6900)$ state and so on have been observed in particular in the $B$ meson decays and in the $e^+e^-$ collisions in various experiments, as summarized in recent reviews [1--5]. The discovery of these exotic QCD states has motivated an extensive theoretical and experimental efforts to understand their properties [6--21]. Notwithstanding, their nature is still largely unknown and further investigations are needed. In the literature, various interpretations of these states, including tightly-bound tetraquark and pentaquark states, loosely-bound meson-meson and baryon-meson molecular states, hadro-charmonium and charmonium-molecule mixtures or the product of rescattering effects, have been proposed (see, for example, Refs. [22--25]).

Among these exotic hadrons, the first charged charmonium-like state $Z_c(3900)$ having charged components $Z_c(3900)^{\pm}$, which was observed in 2013 simultaneously by the BESIII [26] and Belle [27] Collaborations in the
${J/\psi}\pi^{\pm}$ invariant mass distributions from the sequential process $e^+e^- \to Y(4260) \to {\pi^{\mp}}Z_c(3900)^{\pm}\to {\pi^{\mp}}({\pi^{\pm}}J/\psi) \to \pi^+\pi^-{J/\psi}$ at center-of-mass energy of 4.26 GeV, has attracted much attention as the first undoubted tetraquark candidate in the charm sector. The $Z_c(3900)^{\pm} \to {J/\psi}\pi^{\pm}$ decays indicate that the $Z_c(3900)^{\pm}$ are definitely tetraquark states with quark contents $c{\bar c}u{\bar d}$ and $c{\bar c}d{\bar u}$, respectively. Their existence were also confirmed in the analysis of $e^+e^-$ annihilation data taken with the CLEO-c detector at $\sqrt{s}=4.17$ GeV [28] and in the semi-inclusive decays of $b$-flavored hadrons by the $D0$ Collaboration [29]. Evidence for the neutral partner $Z_c(3900)^0$ of the $Z_c(3900)^{\pm}$ and its observation were reported in Refs. [28] and [30], correspondingly
\footnote{$^)$In the literature, the $Z_c(3900)$ is often used for all these three $Z_c(3900)^{\pm}$ and $Z_c(3900)^0$
states.}$^)$ .
The isospin-spin-parity quantum numbers of the $Z_c(3900)$ state were determined by the BESIII Collaboration to be $I(J^P)=1(1^+)$ [31]. It is interesting to note that similar exotic charmonium-like $Z_c(3885)$ states, both charged [32, 33] and neutral [34], were observed in the $(D{\bar D}^*)^{\pm,0}$ mass distributions in the $e^+e^- \to {\pi^{\pm}}(D{\bar D}^*)^{\mp}$ and $e^+e^- \to {\pi^{0}}(D{\bar D}^*)^{0}$ processes. Due to the close values of the masses and widths of the $Z_c(3900)$ and $Z_c(3885)$ states, it is natural to assume that they correspond to the same state [32]. Since the $Z_c(3900)^{\pm}$ states are observed in the $J/\psi{\pi^{\pm}}$ mass distributions slightly above the open-charm $D{\bar D}^*$ mass threshold
\footnote{$^)$Their masses, determined in particular by the BESIII Collaboration as $(3899.0\pm3.6\pm4.9)$ MeV [26], are only $\sim$ 20 MeV above this threshold. It is worthwhile to notice that the charged charmonium-like structures near the $D{\bar D}^*$ and $D^*{\bar D}^*$ mass thresholds were predicted in Ref. [35]. The latter ones, including $Z_c(4020)^{\pm,0}$ and $Z_c(4025)^{\pm,0}$, have been also observed at BESIII in $e^+e^-$ collisions in the experiments
[36, 37] and [38, 39], respectively. To some extent, the charged structures $Z_c(3900)^{\pm}$ and $Z_c(4020/4025)^{\pm}$
are the charmonium analogues of two charged bottomonium-like resonances in the bottomonium sector $Z_b(10610)^{\pm}$ and $Z_b(10650)^{\pm}$ observed by the Belle Collaboration around 10610 MeV and 10650 MeV in the hidden-bottom dipion decays of $\Upsilon(5S)$ [40] whose constituents are $b{\bar b}u{\bar d}$ and $b{\bar b}d{\bar u}$, correspondingly. The measured masses of the $Z_b(10610)^{\pm}$ and $Z_b(10650)^{\pm}$ states are only a few MeV above the thresholds for the open beauty channels $B{\bar B}^*$ and $B^*{\bar B}^*$, which suggests naturally a molecular nature of these states [40].}$^)$, they can be interpreted as $S$-wave $D{\bar D}^*$ resonant or virtual molecular states from the $D{\bar D}^*$ interaction (see, for example, Refs. [41, 42] and those cited below). Also other possible theoretical explanations of the $Z_c(3900)$ states, including the compact tetraquark states [43, 44], hadro-charmonium states [45--47]
\footnote{$^)$In the hadro-charmonium model the resonance $Z_c(3900)$ is a tightly bound color-singlet $c{\bar c}$ core embedded in a light-quark excited hadronic environment.}$^)$
, mixtures of the compact tetraquark states and molecular states [48--50], kinematic threshold cusp effects [51, 52] or kinematical triangle singularities [53--55] have been proposed in recent years (see also references herein below). But, despite a lot of theoretical and experimental efforts, the full understanding of the nature of the intriguing $Z_c(3900)$ states remains a challenging open problem and further studies are also needed here.

To confirm the existence of the exotic $Z_c(3900)^{\pm}$ states, to diagnose them and to set a strong constraints on their properties, it is of importance to investigate their photoproduction on nuclei at energies close to the threshold for their production off a free nucleon. This has the advantage compared to the heavy-ion [56] and pion--nucleus [57] collisions that the interpretation of data from such experiments is less ambiguous owing to a negligible strength of initial-state photon interaction and since the $Z_c(3900)^{\pm}$ production proceeds through a few channels in a cleaner environment - in static cold nuclear matter whose density is sufficiently well known. The study of the $Z_c(3900)^{\pm}$ states photoproduction off a proton target has been carried out previously in Refs. [58--60]. Their exclusive and inclusive production cross sections from pion exchange have been predicted. The discovery potential for charmonium-like states $Z(4430)^{\pm}$ and $Z_c(4200)^{+}$ by their photoproduction on the proton was discussed in Refs. [61] and [62], respectively. The authors of Ref. [63] suggested to search for charged $Z(4430)^{\pm}$ by the nucleon--antinucleon scattering, while the production of neutral $Z(4430)^0$ and $Z_c(4200)^0$ states by antiproton--proton annihilation
was studied in Refs. [64] and [65], correspondingly. It is also worth mentioning that, based on the theoretical predictions obtained in Ref. [58], a search for the $Z_c(3900)^{\pm}$ hadrons through their exclusive photoproduction on the nucleon has been performed by the COMPASS Collaboration [66]. The data cover the range from 7 GeV to 19 GeV in the
center-of-mass energy $W$ of the photon--nucleon system. Unfortunately, no signal of the $Z_c(3900)^{\pm}$ was
observed in the $J/\psi{\pi^{\pm}}$ mass spectrum. A possible explanation of this unexpected result is that the $Z_c(3900)^{\pm}$ are rarely produced and the predicted in [58] total and differential cross sections of ${\gamma}p \to Z_c(3900)^+n$ reaction are overestimated. It is obvious that the particle production off nuclei is enhanced compared to that on the proton. Therefore, searching for the charmonium-like states through their photoproduction on nuclei is an important topic since the photoproduction offers an ideal setup to study them. Such a search could be performed
at the proposed CEBAF upgraded facility with a 22 GeV photon beam, in addition to that of these exotic states in exclusive reactions on the proton target, using the capabilities of the GlueX and CLAS12 detectors for reconstructing $J/\psi \to e^+e^-$ and pions [67, 68].

To motivate it, in this work we present the detailed predictions of some observables for $Z_c(3900)^{\pm}$ production in photonuclear reactions at threshold energies obtained in the framework of the collision model based on the nuclear spectral function within three different scenarios for their internal structures. The predictions can be confronted to the experimental data from the future measurements at the upgraded up to 22 GeV CEBAF facility to discriminate between these scenarios. This would clearly be of great importance as a valuable test of the theoretical predictions in the field of exotic charmonium-like states.

\section*{2. Direct $Z_c(3900)^{\pm}$ photoproduction mechanism: photoproduction cross sections and their ratios}

\hspace{1.5cm} The main goal of this paper is to provide predictions for the $Z_c(3900)^{\pm}$ photoproduction cross sections and their ratios on nuclear targets near the threshold within plausible different scenarios for the
$Z_c(3900)^{\pm}$ internal structures. By near-threshold, we mean the incident photon energy region $E_{\gamma} \le 17.5$ GeV where the $Z_c(3900)^{\pm}$ mesons can be observed in the ${\gamma}p$ and ${\gamma}A$ reactions at an energy-upgraded CEBAF facility [67, 68]
\footnote{$^)$This region corresponds to the relatively "low" excess energies $\epsilon=W-W_{\rm th}$ of the free-space photon--proton system above the $Z_c(3900)^+n$ production threshold $W_{\rm th}=\sqrt{s_{\rm th}}=m_{Z_c}+m_{n}=$ 4.82665 GeV ($m_{Z_c}$ and $m_{n}$ are the $Z_c(3900)^+$ (and $Z_c(3900)^-$) meson and neutron free space masses, respectively ) $0 \le \epsilon \le 1.0$ GeV.}$^)$.
The direct production of the charged components $Z_c(3900)^{\pm}$ on nuclear targets in the laboratory photon energy region of interest can occur in the following ${\gamma}p$ and ${\gamma}n$ elementary processes with zero and one pions in the final states [58--60]
\footnote{$^)$We recall that the free threshold energies, e.g., for the processes (1) and (2) amount, respectively,
to 11.945 and 12.667 GeV.}$^)$:
\begin{equation}
{\gamma}+p \to Z_c(3900)^++n,
\end{equation}
\begin{equation}
{\gamma}+p \to Z_c(3900)^++\pi^-+p,
\end{equation}
\begin{equation}
{\gamma}+p \to Z_c(3900)^++\pi^0+n,
\end{equation}
\begin{equation}
{\gamma}+n \to Z_c(3900)^++\pi^-+n
\end{equation}
and
\begin{equation}
{\gamma}+n \to Z_c(3900)^-+p,
\end{equation}
\begin{equation}
{\gamma}+n \to Z_c(3900)^-+\pi^++n,
\end{equation}
\begin{equation}
{\gamma}+n \to Z_c(3900)^-+\pi^0+p,
\end{equation}
\begin{equation}
{\gamma}+p \to Z_c(3900)^-+\pi^++p.
\end{equation}
The $Z_c(3900)^{\pm}$ mesons, nucleons and pions, produced in these processes, are sufficiently energetic.
Thus, for example, the kinematically allowed $Z_c(3900)^+$ meson and final neutron laboratory momenta in the direct process (1), proceeding on the free target proton at rest, vary within the momentum ranges of 8.347--12.989 GeV/c and 1.011--5.653 GeV/c, respectively, at photon energy of $E_{\gamma}=14$ GeV.
The kinematical characteristics of $Z_c(3900)^{\pm}$ mesons and final nucleons, produced in the reactions (2)--(8),
are similar to those of final particles in the process (1). And outgoing in these reactions pions have an average laboratory momentum $\sim$ 1.0 GeV/c. Since the medium effects are expected to be reduced for high momenta,
we will neglect the medium modifications of the final high-momentum $Z_c(3900)^{\pm}$ hadrons, $\pi$ mesons and nucleons in the case when the reactions (1)--(8) proceed on nucleons embedded in a nuclear target
\footnote{$^)$It should be noted that the behavior of $Z_c(3900)$ tetraquark in a cold nuclear matter has been
studied in the recent work [69]. Accounting for the results for its scalar and vector self-energies obtained in this work, one can expect that even at low energies the in-medium mass shift of the $Z_c(3900)$ tetraquark,
which is determined by the sum [70] of these self-energies, is repulsive and it amounts approximately to 9\%
of its free-space nominal mass for the nuclear matter saturation density $\rho_0$.}$^)$.

Then, neglecting the distortion of the incident photon
\footnote{$^)$It should be pointed out that when the absorption of the incident photons by target nucleons
described by the relevant absorption cross section $\sigma_{{\gamma}N}=$ 0.1 mb is taken into account in Eq. (11),
then the total cross sections (9), (10) for the production of the $Z_c^{\pm}$ mesons on carbon and tungsten target
nuclei are reduced, as our calculations have shown, only by a small fractions about 0.2 and 0.8\%, respectively,
in the considered photon energy range, which cannot change our conclusions, made below, about the possibility
of studying their internal structures in the near-threshold photon-induced reactions.}$^)$
and describing the $Z_c(3900)^{\pm}$ (in what follows denoted as $Z_c^{\pm}$) absorption by intranuclear nucleons by the absorption cross sections $\sigma_{Z_c^{\pm}p}$ and $\sigma_{Z_c^{\pm}n}$, we represent the total cross sections for the production of $Z_c^{\pm}$ mesons on nuclei from the direct photon--induced reactions (1)--(8) as follows [71]:
\begin{equation}
\sigma_{{\gamma}A\to Z_c^+X}^{({\rm dir})}(E_{\gamma})=I_{V}[A,\rho_p,\sigma_{Z_c^+p},\sigma_{Z_c^+n}]
\left[\left<\sigma_{{\gamma}p \to Z_c^+n}(E_{\gamma})\right>_A+\left<\sigma_{{\gamma}p \to Z_c^+{\pi^-}p}(E_{\gamma})\right>_A+
\left<\sigma_{{\gamma}p \to Z_c^+{\pi^0}n}(E_{\gamma})\right>_A\right]+
\end{equation}
$$
+I_{V}[A,\rho_n,\sigma_{Z_c^+p},\sigma_{Z_c^+n}]
\left<\sigma_{{\gamma}n \to Z_c^+{\pi^-}n}(E_{\gamma})\right>_A,
$$
\begin{equation}
\sigma_{{\gamma}A\to Z_c^-X}^{({\rm dir})}(E_{\gamma})=I_{V}[A,\rho_p,\sigma_{Z_c^-p},\sigma_{Z_c^-n}]
\left<\sigma_{{\gamma}p \to Z_c^-{\pi^+}p}(E_{\gamma})\right>_A+
\end{equation}
$$
+I_{V}[A,\rho_n,\sigma_{Z_c^-p},\sigma_{Z_c^-n}]
\left[\left<\sigma_{{\gamma}n \to Z_c^-p}(E_{\gamma})\right>_A+\left<\sigma_{{\gamma}n \to Z_c^-{\pi^+}n}(E_{\gamma})\right>_A+\left<\sigma_{{\gamma}n \to Z_c^-{\pi^0}p}(E_{\gamma})\right>_A\right];
$$
where
\begin{equation}
I_{V}[A,\rho_{p(n)},\sigma_{Z_cp},\sigma_{Z_cn}]=2{\pi}\int\limits_{0}^{R}r_{\bot}dr_{\bot}
\int\limits_{-\sqrt{R^2-r_{\bot}^2}}^{\sqrt{R^2-r_{\bot}^2}}dz
\rho_{p(n)}(\sqrt{r_{\bot}^2+z^2})\times
\end{equation}
$$\times
\exp{\left[-\sigma_{Z_cp}\int\limits_{z}^{\sqrt{R^2-r_{\bot}^2}}\rho_p(\sqrt{r_{\bot}^2+x^2})dx-
\sigma_{Z_cn}\int\limits_{z}^{\sqrt{R^2-r_{\bot}^2}}\rho_n(\sqrt{r_{\bot}^2+x^2})dx
\right]}.
$$
Here, $\rho_p(r)$ and  $\rho_n(r)$ ($r$ is the distance from the nucleus center) are normalized to the numbers of protons $Z$ and neutrons $N$ the local proton and neutron densities of the target nucleus with mass number $A$ ($A=Z+N$), having mass $M_A$ and radius $R$. And $\left<\sigma_{i \to f}(E_{\gamma})\right>_A$
are "in-medium" total cross sections $\sigma_{i \to f}({\sqrt{s^*}})$ ($i \to f={\gamma}p \to Z_c^+n, Z_c^+{\pi^-}p,
Z_c^+{\pi^0}n, Z_c^-{\pi^+}p; {\gamma}n \to Z_c^-p, Z_c^-{\pi^+}n, Z_c^-{\pi^0}p, Z_c^+{\pi^-}n$)
for the production of $Z_c(3900)^{\pm}$ mesons in reactions (1)--(8) at the in-medium ${\gamma}p$ center-of-mass energy $\sqrt{s^*}$
\footnote{$^)$We neglect the difference between proton and neutron masses.}$^)$, averaged over target nucleon binding and Fermi motion. They are defined by Eq. (6) from Ref. [71], in which one needs to make the substitution: ${\gamma}p \to X(3872)p \to i \to f$.

As in Ref. [71], we assume that the "in-medium" cross sections $\sigma_{i \to f}({\sqrt{s^*}})$ for $Z_c^{\pm}$ production in reactions (1)--(8) are equivalent to the vacuum cross sections $\sigma_{i \to f}({\sqrt{s}})$,
in which the center-of-mass energy squared $s$ of the free space photon-proton system for given photon laboratory energy $E_{\gamma}$ and momentum ${\bf p}_{\gamma}$, presented by the formula
\begin{equation}
s=s(E_{\gamma})=W^2=(E_{\gamma}+m_p)^2-{\bf p}_{\gamma}^2=m_p^2+2m_pE_{\gamma},
\end{equation}
is replaced by the in-medium expression
\begin{equation}
  s^*=(E_{\gamma}+E_t)^2-({\bf p}_{\gamma}+{\bf p}_t)^2,
\end{equation}
Here, $E_t$ and ${\bf p}_{t}$ are the total energy and internal momentum of the
struck target nucleons involved in the collision processes (1)--(8). The quantity $E_t$ is determined by Eq. (8)
in Ref. [71].

For the free total cross sections of the photon-induced reaction channels (1)--(8) no data are available
presently at the considered photon energies $E_{\gamma} \le $ 17.5 GeV relevant for the upgraded to the energy of 22 GeV CEBAF facility. Therefore, we have to rely on some theoretical predictions and estimates for them, existing in the
literature at these energies. For the free total cross sections $\sigma_{{\gamma}p \to Z_c^+n}(\sqrt{s})$
and $\sigma_{{\gamma}p \to Z_c^+{\pi^-}p}(\sqrt{s})$ of the reactions (1) and (2) in the considered photon energy range we have used the following parametrizations of the results of calculations of these cross sections here within the charged pion exchange approach [59, 60]
\footnote{$^)$It is also worth mentioning that the approach [59, 60] predicts that the total cross section of the
${\gamma}p \to {Z_c^+}n$ reaction is of about 1.4 nb for the ${\gamma}p$ c.m. energy $W=$ 14 GeV. This value is
comparable with the upper limit for the proton-target $Z_c(3900)$ exclusive photoproduction cross section of $\sim$ 1 nb
measured by the COMPASS Collaboration at an average energy of $<W>=$ 13.8 GeV, once the relevant branching ratio is
taken into account [66]. This fact may give a confidence to us that the predicted in [59, 60] $Z_c$ near-threshold
photoproduction cross sections on a proton target are sufficiently realistic, which implies that the results obtained in
the present work (see below) may be an important tool to provide further insight into its inner structure.}$^)$
:
\begin{equation}
\sigma_{{\gamma}p \to {Z_c^+}n}(\sqrt{s})=42.5\left(1-\frac{s_{\rm th}}{s}\right)^{0.378}
\left(\frac{s_{\rm th}}{s}\right)^{1.357}~[\rm nb],
\end{equation}
\begin{equation}
\sigma_{{\gamma}p \to {Z_c^+}{\pi^-}p}(\sqrt{s})=343.0\left(1-\frac{{\tilde s}_{\rm th}}{s}\right)^{2.122}
\left(\frac{{\tilde s}_{\rm th}}{s}\right)^{3.808}~[\rm nb],
\end{equation}
where
\begin{equation}
s_{\rm th}=W_{\rm th}^2=(m_{Z_c}+m_{n})^2, \,\,
{\tilde s}_{\rm th}={\tilde W}_{\rm th}^2=(m_{Z_c}+m_{\pi^-}+m_{p})^2.
\end{equation}
Here, $m_{\pi^-}$ and $m_p$ are the free space $\pi^-$ meson and proton masses.
According to the predictions of Ref. [60], one may conclude that the total cross sections of the reactions (2) and (3), (2) and (8) are linked by the relations:
\begin{equation}
\sigma_{{\gamma}p \to {Z_c^+}{\pi^0}n}(\sqrt{s})\approx\sigma_{{\gamma}p \to {Z_c^+}{\pi^-}p}(\sqrt{s}),\,\,
\sigma_{{\gamma}p \to {Z_c^-}{\pi^+}p}(\sqrt{s})\approx3\sigma_{{\gamma}p \to {Z_c^+}{\pi^-}p}(\sqrt{s}).
\end{equation}
In line with the "isospin considerations" of the processes (1)--(8), it is natural to assume that the following relations among the total cross sections of the $Z_c(3900)^{\pm}$ production channels (1) and (5), (2) and (6), (3) and (7),
(4) and (8) exist:
\begin{equation}
\sigma_{{\gamma}n \to {Z_c^-}p}(\sqrt{s})\approx\sigma_{{\gamma}p \to {Z_c^+}n}(\sqrt{s}),\,\,
\sigma_{{\gamma}n \to {Z_c^-}{\pi^+}n}(\sqrt{s})\approx\sigma_{{\gamma}p \to {Z_c^+}{\pi^-}p}(\sqrt{s}),
\end{equation}
$$
\sigma_{{\gamma}n \to {Z_c^-}{\pi^0}p}(\sqrt{s})\approx\sigma_{{\gamma}p \to {Z_c^+}{\pi^0}n}(\sqrt{s}),\,\,
\sigma_{{\gamma}n \to {Z_c^+}{\pi^-}n}(\sqrt{s})\approx\sigma_{{\gamma}p \to {Z_c^-}{\pi^+}p}(\sqrt{s}).
$$
It is worth noting that for photon energies near the thresholds ($W$ $\sim$ 6.0 GeV), the cross sections for the $Z_c(3900)^{+}$ and $Z_c(3900)^{-}$ production in ${\gamma}p$ collisions are predicted to be of the order of 20--30 nanobarns [59, 60], which are well within reach of the upgraded up to 22 GeV CEBAF facility.
We will use the relations (14)--(18) in our subsequent calculations as a guideline for
a reasonable estimation of the $Z_c(3900)^{\pm}$ yield in ${\gamma}A$ reactions.

The local proton and neutron densities, adopted in the calculations of the quantities\\
$I_{V}[A,\rho_{p(n)},\sigma_{Z_cp},\sigma_{Z_cn}]$ entering into Eqs. (9) and (10), for the target nuclei $^{12}_{6}$C, $^{27}_{13}$Al, $^{40}_{20}$Ca, $^{63}_{29}$Cu, $^{93}_{41}$Nb, $^{112}_{50}$Sn, $^{184}_{74}$W, $^{208}_{82}$Pb and $^{238}_{92}$U considered in the present work are given in Ref. [71]. As before in Ref. [71], for medium-weight $^{93}_{41}$Nb, $^{112}_{50}$Sn and heavy $^{184}_{74}$W, $^{208}_{82}$Pb, $^{238}_{92}$U target nuclei we use the neutron density $\rho_n(r)$ in the 'skin' form.

To estimate the rates of the $Z_c^{\pm}$ photoproduction off nuclei in different scenarios for their intrinsic structures, we have to specify the effective input $Z_c^{\pm}$--nucleon absorption cross sections $\sigma_{Z_c^{\pm}p}$ and $\sigma_{Z_c^{\pm}n}$ in these scenarios. They determine the numbers (11) of the target nucleons participating in the direct processes (1)--(8). To this end in our exploratory study we consider three different popular scenarios for the $Z_c^{\pm}$ [72]: i) compact, $\sim$ 1 fm, diquark-antidiquark tetraquark bound states: relatively tightly bound pairs $[cu]$ and $[{\bar c}{\bar d}]$ for the $Z_c^+$ and $[cd]$ and $[{\bar c}{\bar u}]$ ones for the $Z_c^-$, which interact by the gluonic color force [43, 44, 56, 69, 73--81], ii) genuine resonant or virtual bound
$D{\bar D}^*+{\bar D}D^*$ molecular states - $S$-wave hadronic molecules formed by pairs of charmed and anticharmed mesons [41, 82--101]
\footnote{$^)$The observed $Z_c(3900)$ mass  ($m_{Z_c}=3887.1\pm2.6$ MeV [102]), obtained by averaging those from
experimental analyses of both charged and neutral $Z_c(3900)$, lies about 10.6 and 12 MeV above the thresholds of $D^+{\bar D}^{*0}$ and $D^{*+}{\bar D}^0$ for the $Z_c^+$ and of $D^-D^{*0}$ and $D^{*-}D^0$ for the $Z_c^-$, respectively. This makes the $Z_c(3900)$ a good candidate for either a $D{\bar D}^*$ virtual state below the lowest threshold or an above-threshold molecular resonance (see, for example, Refs. [41, 93, 100, 103]).}$^)$,
and iii) hybrid states - a states in which the $Z_c^{\pm}$ are considered as the mixtures of
compact diquark-antidiquark and molecular components [48--50].

Since the mass of the observed charmonium-like structure $Z_c(3900)^{\pm}$ is very close to that of the famous
$X(3872)$, it can be considered as the charged isovector partner of the $X(3872)$ state [43]. Therefore,
interpreting the $Z_c(3900)^{\pm}$ as a compact tetraquarks, it is natural to assume for the high-momentum $Z_c(3900)^{\pm}$--proton (neutron) absorption cross sections $\sigma_{{Z_c^{\pm}}p(n)}^{{\rm 4{q}}}$ in this picture for the $Z_c(3900)^{\pm}$ the same value as that for the absorption cross section of the $X(3872)$ resonance with similar momenta considering it as a compact tetraquark as well, i.e., $\sigma_{{Z_c^{\pm}}p(n)}^{{\rm 4{q}}}=13.3$ mb [71].

Within the "pure" hadronic molecular interpretation of the $Z_c(3900)^{\pm}$, their wave functions take the
following charge-conjugation forms [41, 81, 83, 85, 89]:
\begin{equation}
|Z_c(3900)^+>_{\rm mol}=\frac{1}{\sqrt{2}}\left(|D^+{\bar D}^{*0}>+|D^{*+}{\bar D}^0>\right),
\end{equation}
\begin{equation}
|Z_c(3900)^->_{\rm mol}=\frac{1}{\sqrt{2}}\left(|D^-D^{*0}>+|D^{*-}D^0>\right).
\end{equation}
In lieu of an appropriate theory, we assume that the $Z_c(3900)^{\pm}$ states dissociate while one of their constituent charmed mesons flying with an average laboratory momentum $\sim$ 5 GeV/c (see above) scatters (elastically or inelastically) with the intranuclear nucleon (proton or neutron) and the other charmed meson is a spectator (cf. [71, 104, 105])
\footnote{$^)$It should be pointed out that this assumption is sufficiently well justified only when the average distance between the constituents in the hadronic molecule is much larger than a typical radius of strong interactions (a few fermi). In this case the constituents can be considered as on-shell individual particles flying together and interactions between them can be ignored [104, 105]. But in our present case of $Z_c(3900)^{\pm}$, a relatively high excitation energy $\sim$ 10 MeV over the threshold pushes this picture to its limit of validity. Thus, using the expression $r_{Z_c}=1/\sqrt{4\mu_0|\delta_{Z_c}|}$ for the r.m.s. size $r_{Z_c}$ of the intermeson distance [46, 71], where $\mu_0$ is the $D{\bar D}^*$ (${\bar D}D^*$) reduced mass and $\delta_{Z_c}$ is "the binding energy" $\delta_{Z_c}=m_{D}+m_{{\bar D}^*}-m_{Z_c}$ $(\delta_{Z_c}=m_{{\bar D}}+m_{D^*}-m_{Z_c})$ of the molecule, and taking $\delta_{Z_c}=m_{{\bar D}^0}+m_{D^{*+}}-m_{Z_c}=-12$ MeV, we find that the charmed mesons ${\bar D}^0$ and $D^{*+}$ in the $Z_c(3900)^+$ have an insignificant r.m.s. separation: $r_{Z_c}=0.92$ fm. This size scale is small and it is close to the range, where the charmed mesons as a spatially extended objects with mean spatial sizes $\sim$ 0.4 fm [106] start to overlap and possibly can not be considered as individual particles. This implies that the $Z_c(3900)^{\pm}$ can not be reliable described as molecular states [43, 46, 75] and it may not be correct to treat their interactions with target nucleons in the manner adopted in the present work. The situation could be improved only at the price of making the sizes of open charm mesons
much smaller than what they are assumed to be. But in view of the absence of such possibility and a reliable theory, for an estimate of the $Z_c(3900)^{\pm}N$ absorption cross sections in molecular scenario, we will nevertheless follow this treatment.}$^)$.
Then, assuming that the total cross sections of the free high-momentum ${\bar D}^{*0}N$, $D^{*0}N$ and $D^{*+}N$,  $D^{*-}N$ interactions are the same as those for the ${\bar D}^{0}N$, $D^{0}N$ and $D^{+}N$,  $D^{-}N$
ones [107--110], we can evaluate the cross sections for $Z_c(3900)^{\pm}$ absorption in the molecular scenario, $\sigma_{{Z_c^{\pm}}p(n)}^{\rm mol}$, as [71, 107--110]:
\begin{equation}
\sigma_{{Z_c^+}p}^{\rm mol}\approx\sigma_{D^+p}^{\rm el}+\sigma_{D^+p}^{\rm in}+
\sigma_{{\bar D}^0p}^{\rm el}+\sigma_{{\bar D}^0p}^{\rm in},
\end{equation}
\begin{equation}
\sigma_{{Z_c^+}n}^{\rm mol}\approx\sigma_{D^+n}^{\rm el}+\sigma_{D^+n}^{\rm in}+
\sigma_{{\bar D}^0n}^{\rm el}+\sigma_{{\bar D}^0n}^{\rm in}+\sigma_{{\bar D}^0n \to D^-p}+
\sigma_{D^+n \to D^0p};
\end{equation}
\begin{equation}
\sigma_{{Z_c^-}p}^{\rm mol}\approx\sigma_{D^-p}^{\rm el}+\sigma_{D^-p}^{\rm in}+
\sigma_{D^0p}^{\rm el}+\sigma_{D^0p}^{\rm in}+\sigma_{D^-p \to {\bar D}^0n}+\sigma_{D^0p \to D^+n},
\end{equation}
\begin{equation}
\sigma_{{Z_c^-}n}^{\rm mol}\approx\sigma_{D^0n}^{\rm el}+\sigma_{D^0n}^{\rm in}+\sigma_{D^-n}^{\rm el}+
\sigma_{D^-n}^{\rm in}.
\end{equation}
Here, $\sigma_{D^+p(n)}^{\rm el(in)}$ and $\sigma_{D^-p(n)}^{\rm el(in)}$ as well as $\sigma_{{\bar D}^0p(n)}^{\rm el(in)}$ and $\sigma_{D^0p(n)}^{\rm el(in)}$ are the elastic (inelastic) cross sections of the free
$D^+p$($D^+n$) and $D^-p$($D^-n$) as well as ${\bar D}^0p$(${\bar D}^0n$) and $D^0p$($D^0n$)
interactions, respectively. And, $\sigma_{{\bar D}^0n \to D^-p}$, $\sigma_{D^+n \to D^0p}$,
$\sigma_{D^-p \to {\bar D}^0n}$ and $\sigma_{D^0p \to D^+n}$ are the total cross sections of the free charge-exchange
reactions ${\bar D}^0n \to D^-p$, $D^+n \to D^0p$, $D^-p \to {\bar D}^0n$ and $D^0p \to D^+n$, correspondingly.
In our calculations we adopt for them the following constants which are relevant to the momentum regime above of 1 GeV/c of interest:
$\sigma_{D^+p(n)}^{\rm el}=\sigma_{D^-p(n)}^{\rm el}=\sigma_{D^0p(n)}^{\rm el}=\sigma_{{\bar D}^0p(n)}^{\rm el}=10$ mb,
$\sigma_{D^+p(n)}^{\rm in}=\sigma_{D^0p(n)}^{\rm in}=10$ mb, $\sigma_{D^-p(n)}^{\rm in}=\sigma_{{\bar D}^0p(n)}^{\rm in}=0$, $\sigma_{{\bar D}^0n \to D^-p}=\sigma_{D^+n \to D^0p}=\sigma_{D^-p \to {\bar D}^0n}=\sigma_{D^0p \to D^+n}=12$ mb [107--109]. Using these values, we obtain that
$\sigma_{{Z_c^+}p}^{\rm mol}=30$ mb, $\sigma_{{Z_c^+}n}^{\rm mol}=54$ mb and
$\sigma_{{Z_c^-}p}^{\rm mol}=54$ mb, $\sigma_{{Z_c^-}n}^{\rm mol}=30$ mb.

In the hybrid scenario, it is assumed that the $Z_c(3900)^{\pm}$ wave functions are a linear superpositions of the compact four-quark (or 4$q$) and molecular components [48--50]:
\begin{equation}
|Z_c(3900)^+>_{\rm hyb}=\alpha|c{\bar c}u{\bar d}>+\frac{\beta}{\sqrt{2}}\left(|D^+{\bar D}^{*0}>+|D^{*+}{\bar D}^0>\right),
\end{equation}
\begin{equation}
|Z_c(3900)^->_{\rm hyb}=\alpha|c{\bar c}{\bar u}d>+\frac{\beta}{\sqrt{2}}\left(|D^-D^{*0}>+|D^{*-}D^0>\right).
\end{equation}
Here, the amplitudes $\alpha$ and $\beta$ squared, $\alpha^2$ and $\beta^2$, represent the probabilities to find a relevant "elementary" compact non-molecular and two-body molecular hadronic components in the $Z_c(3900)^{\pm}$ resonances, respectively. They are normalized as
\begin{equation}
\alpha^2+\beta^2=1
\end{equation}
and are named in the literature [48, 50, 111--113], correspondingly, as the "elementariness" (or the "elementarity" [112, 113]) $Z$ and "compositeness" $X$
\footnote{$^)$It is worthwhile to note that the quantity ${\bar X}_A$, which can be used to estimate the compositeness for bound states, resonances and virtual states knowing the scattering length and the effective range of the respective two-body channel, was introduced in Ref. [111]. A recent review article that summarizes the current understanding of the compositeness is presented in Ref. [114].}$^)$.
The limiting case of $\alpha^2=0$, $\beta^2=1$ corresponds to the pure molecular interpretation of the $Z_c(3900)^{\pm}$ states, while the case of $\alpha^2=1$, $\beta^2=0$ refers to their pure compact four-quark treatment.
Recently, in Ref. [48] the $Z_c(3900)$ state was studied as the ${J/\psi}\pi$ and $D{\bar D}^*$ coupled-channel system and in particular the compositeness ${\bar X}_A$ of the $D{\bar D}^*$ component in this state was calculated to be less
than 0.5 (${\bar X}_A\approx0.36$), indicating that other hadronic components or components of short-distance nature compared to the $D{\bar D}^*$ - compact quark states could be also important in the formation of the $Z_c(3900)$.
The value of ${\bar X}_A$ of about 0.5 (${\bar X}_A\approx0.48$) for the $Z_c(3900)$ was obtained very recently
in Ref. [50] in the unified analysis of the experimental data of the $\pi^+\pi^-$ and ${J/\psi}\pi^{\pm}$ invariant
mass spectra for $e^+e^- \to {J/\psi}\pi^+\pi^-$ and the $D^0D^{*-}$ mass spectrum for $e^+e^- \to {D^0}D^{*-}\pi^+$ at $e^+e^-$ center-of-mass energies of 4.23 and 4.26 GeV, suggesting that the $D{\bar D}^*$ molecular and non-molecular components are of similar importance in the structure of the $Z_c(3900)$.
In view of the above, following Refs. [49, 50], we will assume that the $Z_c(3900)^{\pm}$ wave functions (25), (26) contain 50\% of the genuine non-molecular 4$q$ component and 50\% of the molecular $D{\bar D}^*$+ ${\bar D}D^*$
component ($\alpha^2=\beta^2=0.5$). In the hybrid interpretation (25), (26) of the $Z_c(3900)^{\pm}$, we can represent  the $Z_c(3900)^{\pm}$--proton (neutron) absorption cross sections $\sigma_{{Z_c^{\pm}}p(n)}^{\rm hyb}$ in the following incoherent probability-weighted sum [71]
\footnote{$^)$The role of the interference effects between the compact tetraquark and molecular contributions to the
$Z_c(3900)$--nucleon absorption cross sections $\sigma_{Z_c^{\pm}p(n)}^{\rm hyb}$ (28) in the hybrid pictures
(25), (26) of the $Z_c^{\pm}$ is expected to be insignificant due to the fact that the mechanisms of their dissociation when they interact with the surrounding nucleons are quite different in the molecular and non-molecular configurations: loosely-bound charm meson molecules $D{\bar D}^*$ are destroyed via the collisions of the constituent $D$ and
${\bar D}^*$ mesons with the intranuclear nucleons, while the tightly-bound compact tetraquarks having small spatial size are disappeared, as one might assume, when they collide as a single "elementary" objects with these nucleons.
And in these collisions the final states, as can be expected, would be different. }$^)$
:
\begin{equation}
\sigma_{Z_c^{\pm}p(n)}^{\rm hyb}=\alpha^2\sigma_{Z_c^{\pm}p(n)}^{\rm 4{q}}+\beta^2\sigma_{Z_c^{\pm}p(n)}^{\rm mol}.
\end{equation}
According to the above, we set $\sigma_{Z_c^{\pm}p(n)}^{\rm 4{q}}=13.3$ mb and $\sigma_{Z_c^+p(n)}^{\rm mol}=30~(54)$
mb, $\sigma_{Z_c^-p(n)}^{\rm mol}=54~(30)$ mb. With these values, the $\sigma_{Z_c^{\pm}p(n)}^{\rm hyb}$ absorption cross sections (28) are $\sigma_{Z_c^{+}p(n)}^{\rm hyb}=21.65~(33.65)$ mb and $\sigma_{Z_c^{-}p(n)}^{\rm hyb}=33.65~(21.65)$ mb for the non-molecular and molecular probabilities of the $Z_c(3900)^{\pm}$ 50\% and 50\%. Now, we summarize here the results obtained above for the $Z_c(3900)^{\pm}$--proton (neutron) absorption cross sections $\sigma_{{Z_c^{\pm}}p(n)}$ in the adopted scenarios for the $Z_c(3900)^{\pm}$:
\begin{equation}
\sigma_{{Z_c^+}p}=\left\{
\begin{array}{lll}
	13.3~{\rm mb}
	&\mbox{for tetraquark (4q) state}, \\
	&\\
    21.65~{\rm mb}
	&\mbox{for hybrid state (50\%,50\%)},\\
	&\\
    30~{\rm mb}
	&\mbox{for $D{\bar D}^*$ molecule};
\end{array}
\right.	
\end{equation}
\begin{equation}
\sigma_{{Z_c^+}n}=\left\{
\begin{array}{lll}
	13.3~{\rm mb}
	&\mbox{for tetraquark (4q) state}, \\
	&\\
    33.65~{\rm mb}
	&\mbox{for hybrid state (50\%,50\%)},\\
	&\\
    54~{\rm mb}
	&\mbox{for $D{\bar D}^*$ molecule}
\end{array}
\right.	
\end{equation}
and
\begin{equation}
\sigma_{{Z_c^-}p}=\sigma_{{Z_c^+}n},\,\,\,\,\sigma_{{Z_c^-}n}=\sigma_{{Z_c^+}p}.
\end{equation}
One can see that the $Z_c(3900)^{\pm}$ as $D{\bar D}^*$ molecules have the largest absorption cross sections and,
hence, they are expected to be more easily absorbed in this case than their other internal configurations in a nuclear medium. We will employ these values for the quantities $\sigma_{Z_c^{\pm}p(n)}$ in our subsequent total and differential cross-section calculations. It is also worth noting that according to Eqs. (9)--(11), (29)--(31) and our assumptions
about the elementary $Z_c(3900)^{\pm}$ production cross sections in ${\gamma}N$ collisions for nuclei, like $^{12}_{6}$C, $^{40}_{20}$Ca, with the same local proton $\rho_p(r)$ and neutron $\rho_n(r)$ densities the $Z_c(3900)^+$ and $Z_c(3900)^-$ total and differential (defined below) production cross sections are the same.
\begin{figure}[!h]
\begin{center}
\includegraphics[width=15.0cm]{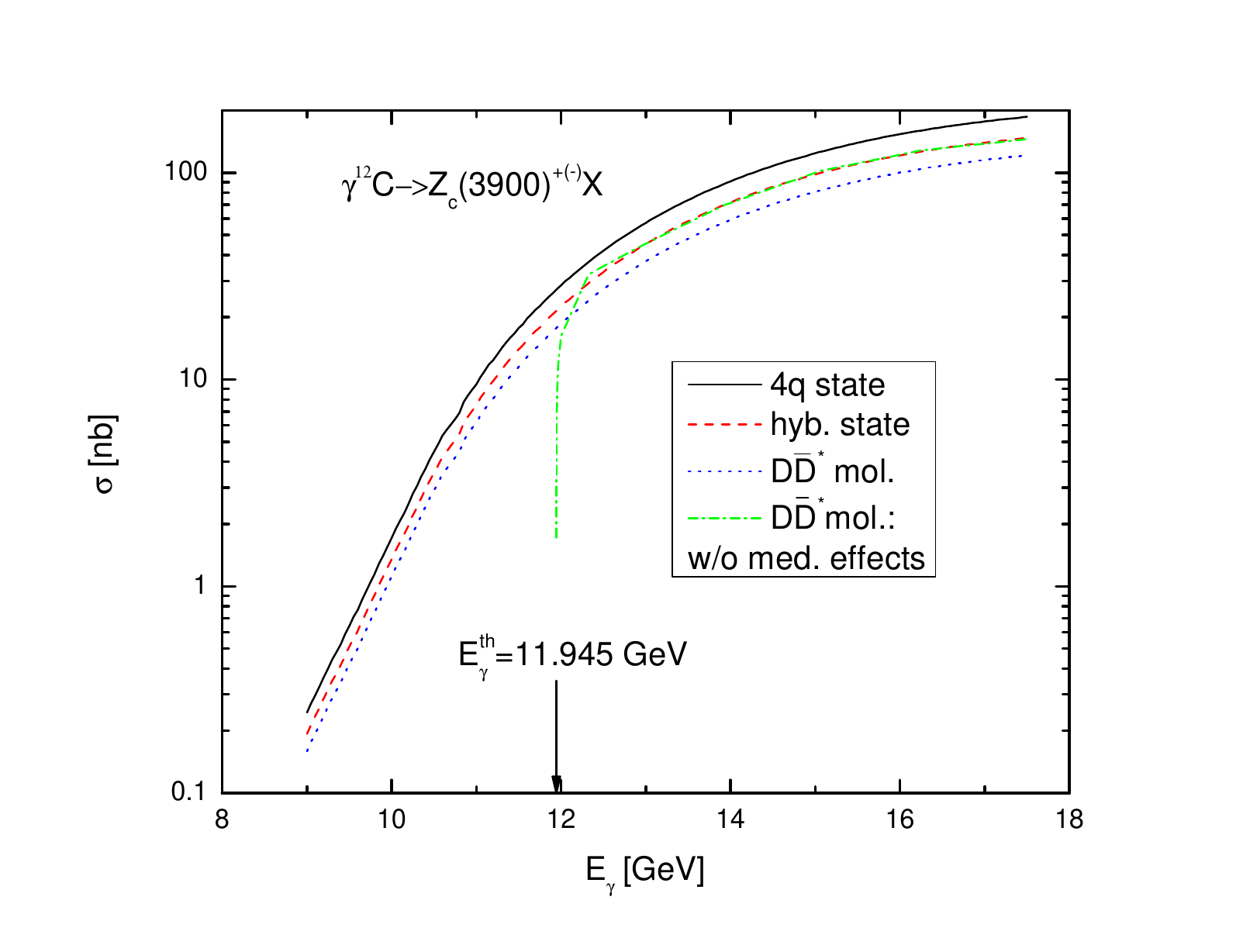}
\vspace*{-2mm} \caption{(Color online.) Excitation functions for the production of $Z_c(3900)^+$
and $Z_c(3900)^-$ mesons off $^{12}$C , respectively, from the direct processes (1)--(4) and (5)--(8)
proceeding on an off-shell target nucleons and on a free ones being at rest. The curves are calculations in
the scenarios, in which the $Z_c(3900)^{\pm}$ are treated as a purely compact four quark (4$q$) states, as a purely
molecular states, or as a hybrid states: mixtures of the non-molecular (compact) and molecular (non-compact) components,
in which there are 50\% of the 4$q$ components and 50\% molecular components.
The arrow indicates the threshold energy for the $Z_c(3900)^+$ and $Z_c(3900)^-$
photoproduction on a free nucleons in reactions (1) and (5), correspondingly.}
\label{void}
\end{center}
\end{figure}

We consider now two additional integral observables that can be used to verify properties of the $Z_c(3900)^{\pm}$
tetraquark states. Both are cross-section ratios, which are sensitive to the $Z_c(3900)^{\pm}$--nucleon absorption cross sections (and, hence, to their intrinsic structures) and which are more robust than cross sections themselves to systematic effects associated with luminosity and detector efficiency. The first observable - the so-called $Z_c(3900)^{\pm}$ transparency ratio - is the ratio of the nuclear $Z_c(3900)^{\pm}$ photoproduction cross sections (9), (10) divided by $A$ times the same quantities on a free nucleon (cf. Ref. [71] and references herein):
\begin{equation}
S_A=\frac{\sigma_{{\gamma}A \to Z_c^{\pm}X}^{({\rm dir})}(E_{\gamma})}{A~\sigma_{{\gamma}N \to Z_c^{\pm}N^{\prime}}(\sqrt{s(E_{\gamma})})},
\end{equation}
where
\begin{equation}
\sigma_{{\gamma}N \to Z_c^{+}N^{\prime}}(\sqrt{s(E_{\gamma})})=\sigma_{{\gamma}p \to Z_c^{+}n}(\sqrt{s(E_{\gamma})})
\end{equation}
and
\begin{equation}
\sigma_{{\gamma}N \to Z_c^{-}N^{\prime}}(\sqrt{s(E_{\gamma})})=\sigma_{{\gamma}n \to Z_c^{-}p}(\sqrt{s(E_{\gamma})}).
\end{equation}
The second observable is the $Z_c(3900)^{\pm}$ transparency ratio $S_A$ normalized to a light nucleus like $^{12}$C [71, 115]:
\begin{equation}
T_A=\frac{S_A}{S_C}=\frac{12}{A}\frac{\sigma_{{\gamma}A \to Z_c^{\pm}X}^{({\rm dir})}(E_{\gamma})}
{\sigma_{{\gamma}C \to Z_c^{\pm}X}^{({\rm dir})}(E_{\gamma})}.
\end{equation}
Evidently that the definition of the quantities $S_A$ and $T_A$ via Eqs. (32) and (35) implies that they
should be considered only at above threshold photon energies. But since the right-hand side of Eq. (35) is defined
both above and below the threshold, we will employ it in calculating the transparency ratio $T_A$ also at subthreshold
photon energies (cf. Fig. 9 given below). The formulas (32)--(35) will be used in our calculations of the integral observables $S_A$ and $T_A$.
\begin{figure}[!h]
\begin{center}
\includegraphics[width=15.0cm]{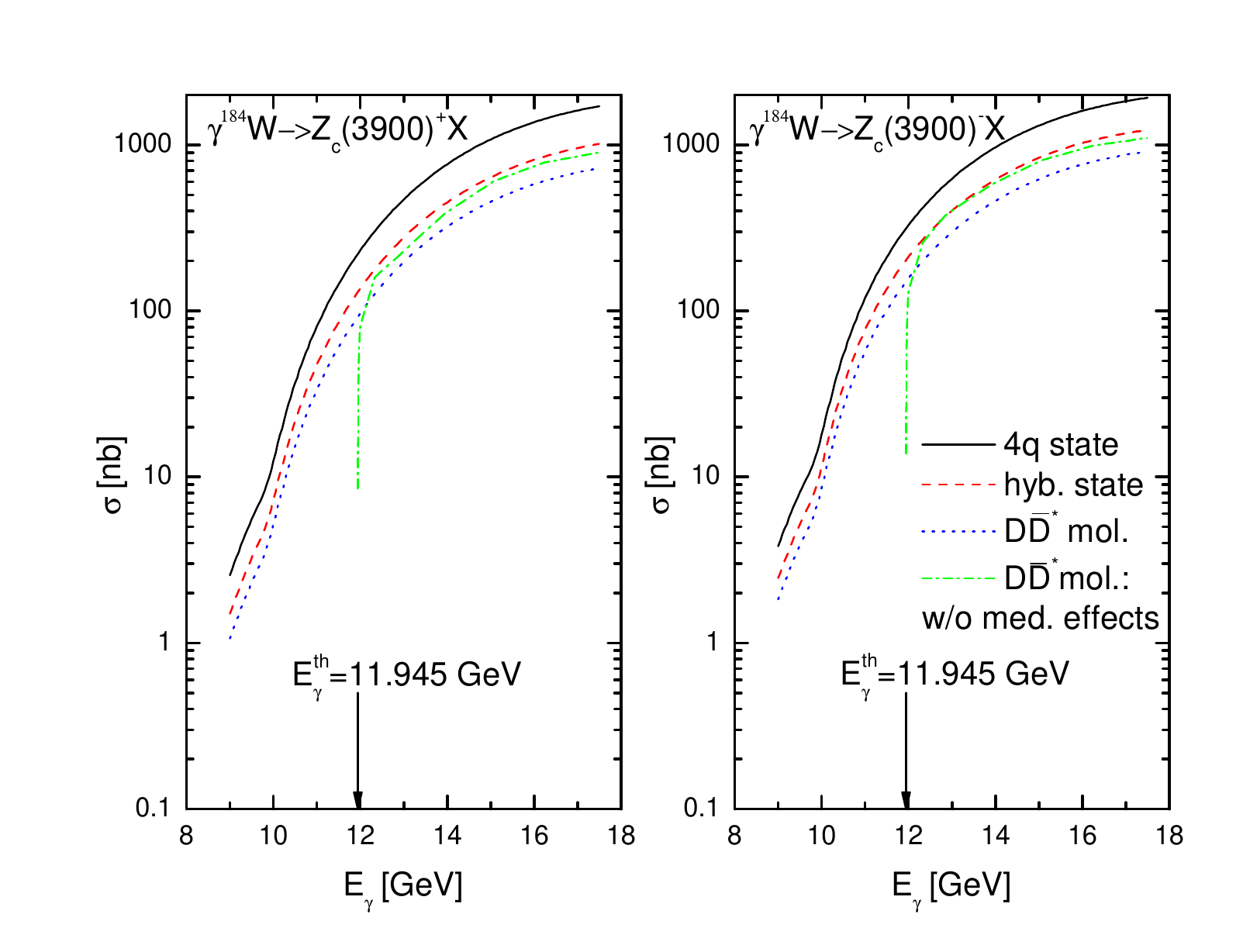}
\vspace*{-2mm} \caption{(Color online.) Excitation functions for the production of $Z_c(3900)^+$ (left panel)
and $Z_c(3900)^-$ (right panel) mesons off $^{184}$W , respectively, from the direct processes (1)--(4) and (5)--(8)
proceeding on an off-shell target nucleons and on a free ones being at rest. The curves are calculations in
the scenarios, in which the $Z_c(3900)^{\pm}$ are treated as a purely compact four quark (4$q$) states, as a purely
molecular states, or as a hybrid states: mixtures of the non-molecular (compact) and molecular (non-compact) components,
in which there are 50\% of the 4$q$ components and 50\% molecular components.
The arrows indicate the threshold energy for the $Z_c(3900)^+$ and $Z_c(3900)^-$
photoproduction on a free nucleons in reactions (1) and (5), correspondingly.}
\label{void}
\end{center}
\end{figure}

The information on the $Z_c(3900)^{\pm}$ inner structure can also be extracted from the comparison of the measured and calculated momentum distributions of $Z_c(3900)^{\pm}$ mesons from nuclei in the photon energy range of interest.
Therefore, we consider now the momentum-dependent inclusive differential cross sections for their production
with momentum $p_{Z_ c^{\pm}}$ on nuclei from the photon-induced elementary production channels (1)--(8).
Accounting for the fact that the $Z_c(3900)^{\pm}$ mesons move in the nucleus essentially forward in the lab system
\footnote{$^)$Thus, for example, the maximum angle of their production on a free target nucleon at rest in
reactions (1), (5) is about 4.5$^{\circ}$ at photon energy of 14 GeV.}$^)$,
we will calculate the $Z_c(3900)^{\pm}$ momentum distributions from considered target nuclei
for the laboratory solid angles ${\Delta}{\bf \Omega}_{Z_c^{\pm}}$ = $0^{\circ} \le \theta_{Z_c^{\pm}} \le 10^{\circ}$,
and $0 \le \varphi_{Z_c^{\pm}} \le 2{\pi}$. Then, according to Eqs. (9)--(11) and Ref. [71], we represent these distributions as follows:
\begin{equation}
\frac{d\sigma_{{\gamma}A\to {Z_c^+}X}^{({\rm dir})}
(p_{\gamma},p_{Z_c^+})}{dp_{Z_c^+}}=
2{\pi}I_{V}[A,\rho_p,\sigma_{Z_c^+p},\sigma_{Z_c^+n}]
\int\limits_{\cos10^{\circ}}^{1}d\cos{{\theta_{Z_c^+}}}\times
\end{equation}
$$\times
\left[\left<\frac{d\sigma_{{\gamma}p\to {Z_c^+}{n}}(p_{\gamma},
p_{Z_c^+},\theta_{Z_c^+})}{dp_{Z_c^+}d{\bf \Omega}_{Z_c^+}}\right>_A+
\left<\frac{d\sigma_{{\gamma}p\to {Z_c^+}{\pi^-}p}(p_{\gamma},
p_{Z_c^+},\theta_{Z_c^+})}{dp_{Z_c^+}d{\bf \Omega}_{Z_c^+}}\right>_A+
\left<\frac{d\sigma_{{\gamma}p\to {Z_c^+}\pi^0{n}}(p_{\gamma},
p_{Z_c^+},\theta_{Z_c^+})}{dp_{Z_c^+}d{\bf \Omega}_{Z_c^+}}\right>_A\right]+
$$
$$
+2{\pi}I_{V}[A,\rho_n,\sigma_{Z_c^+p},\sigma_{Z_c^+n}]
\int\limits_{\cos10^{\circ}}^{1}d\cos{{\theta_{Z_c^+}}}
\left<\frac{d\sigma_{{\gamma}n\to {Z_c^+}\pi^-{n}}(p_{\gamma},
p_{Z_c^+},\theta_{Z_c^+})}{dp_{Z_c^+}d{\bf \Omega}_{Z_c^+}}\right>_A,
$$
\begin{equation}
\frac{d\sigma_{{\gamma}A\to {Z_c^-}X}^{({\rm dir})}
(p_{\gamma},p_{Z_c^-})}{dp_{Z_c^-}}=
2{\pi}I_{V}[A,\rho_p,\sigma_{Z_c^-p},\sigma_{Z_c^-n}]
\int\limits_{\cos10^{\circ}}^{1}d\cos{{\theta_{Z_c^-}}}
\left<\frac{d\sigma_{{\gamma}p\to {Z_c^-}\pi^+{p}}(p_{\gamma},
p_{Z_c^-},\theta_{Z_c^-})}{dp_{Z_c^-}d{\bf \Omega}_{Z_c^-}}\right>_A+
\end{equation}
$$
+2{\pi}I_{V}[A,\rho_n,\sigma_{Z_c^-p},\sigma_{Z_c^-n}]
\int\limits_{\cos10^{\circ}}^{1}d\cos{{\theta_{Z_c^-}}}\times
$$
$$\times
\left[\left<\frac{d\sigma_{{\gamma}n\to {Z_c^-}{p}}(p_{\gamma},
p_{Z_c^-},\theta_{Z_c^-})}{dp_{Z_c^-}d{\bf \Omega}_{Z_c^-}}\right>_A+
\left<\frac{d\sigma_{{\gamma}n\to {Z_c^-}{\pi^+}n}(p_{\gamma},
p_{Z_c^-},\theta_{Z_c^-})}{dp_{Z_c^-}d{\bf \Omega}_{Z_c^-}}\right>_A+
\left<\frac{d\sigma_{{\gamma}n\to {Z_c^-}\pi^0{p}}(p_{\gamma},
p_{Z_c^-},\theta_{Z_c^-})}{dp_{Z_c^-}d{\bf \Omega}_{Z_c^-}}\right>_A\right].
$$
Here,
$\left<\frac{d\sigma_{i \to f}(p_{\gamma},
p_{Z_c^{\pm}},\theta_{Z_c^{\pm}})}{dp_{Z_c^{\pm}}d{\bf \Omega}_{Z_c^{\pm}}}\right>_A$
are the off-shell differential cross sections for production of $Z_c^{\pm}$ mesons
with momenta ${\bf p}_{Z_c^{\pm}}$ in the processes (1)--(8) ($i \to f$),
averaged over the Fermi motion and binding energy of the intranuclear nucleons.
Such differential cross sections for production of $Z_c^{+}$ and $Z_c^-$ mesons
in the processes (1) and (5) with two-body final states can be expressed by Eqs. (28), (31)--(39)
from Ref. [116], in which one needs to make the
substitution: $({\gamma}p \to \Upsilon(1S)p) \to ({\gamma}p \to Z_c^+n)$, $\Upsilon(1S) \to Z_c^+$ and
$({\gamma}p \to \Upsilon(1S)p) \to ({\gamma}n \to Z_c^-p)$, $\Upsilon(1S) \to Z_c^-$.
In order to calculate the c.m. $Z_c^{\pm}$ angular distributions in reactions (1) and (5)
(cf. Eq. (34) from Ref. [116]) one needs to know their exponential $t$-slope parameters $b_{Z_c^{\pm}}$ in
the threshold energy region. In view of the above, we adopt for both these parameters the value of
2.0 GeV$^{-2}$ obtained in Ref. [71] for the $X(3872)$ slope parameter $b_{X(3872)}$ for incident photon
energy of 13 GeV. The analogous differential cross sections of the precesses (2)--(4) and (6)--(8) with
three-body final states, entering into Eqs. (36), (37), can be described by the expressions (6), (28)--(31)
from Ref. [117] in which the respective changes have to be made. For the sake of conciseness, we do not give these expressions here as well.
\begin{figure}[!h]
\begin{center}
\includegraphics[width=15.0cm]{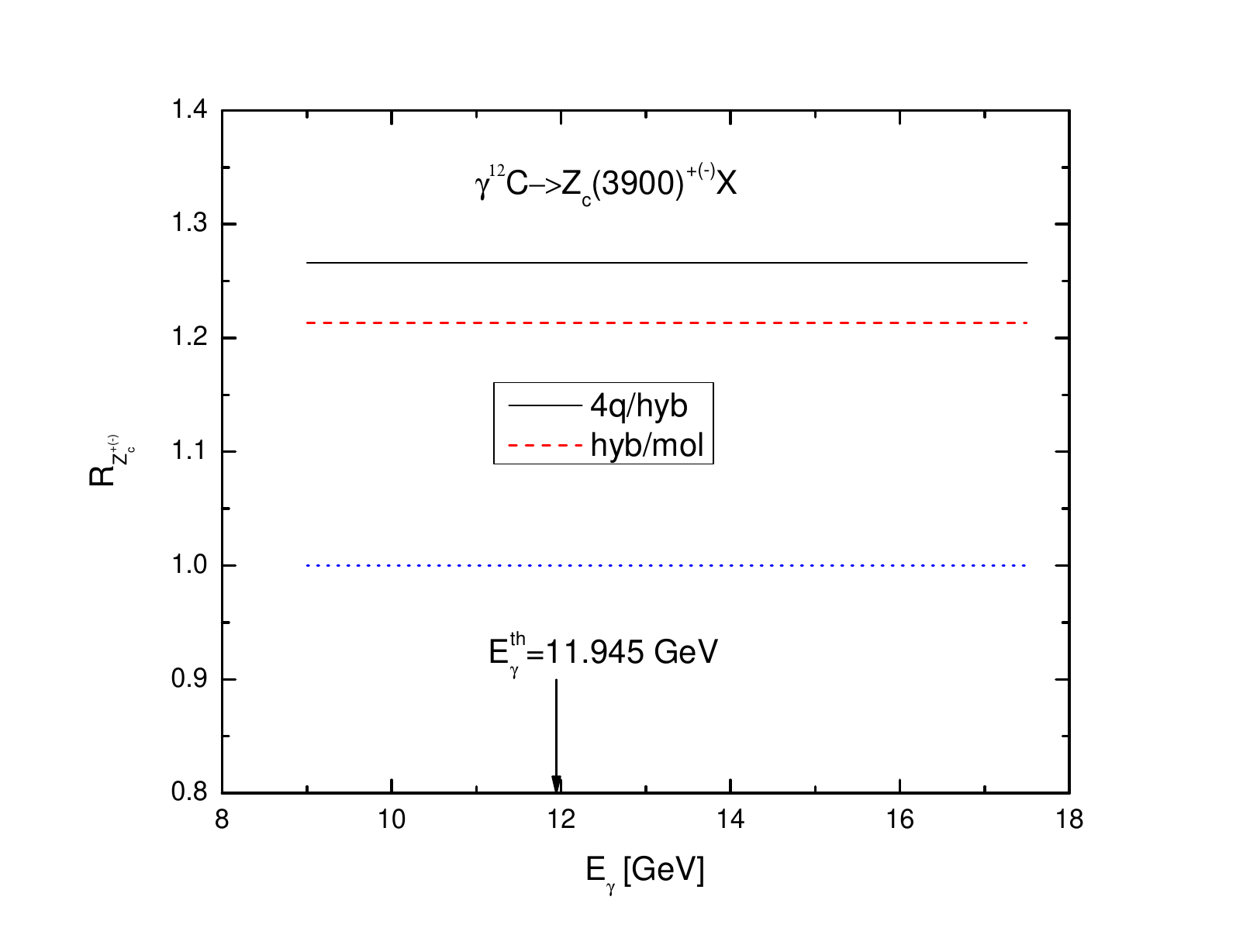}
\vspace*{-2mm} \caption{(Color online.) Ratios between the $Z_c(3900)^{+}$ and between the $Z_c(3900)^-$ production cross sections on $^{12}$C, shown in Fig. 1 and calculated in the compact tetraquark and hybrid scenarios, in the hybrid and molecular scenarios of the $Z_c(3900)^{\pm}$ for an off-shell target nucleons, as functions of photon energy. The arrow indicates the threshold energy for the $Z_c(3900)^+$ and $Z_c(3900)^-$ photoproduction on a free nucleons in reactions (1) and (5), respectively.}
\label{void}
\end{center}
\end{figure}
\begin{figure}[!h]
\begin{center}
\includegraphics[width=15.0cm]{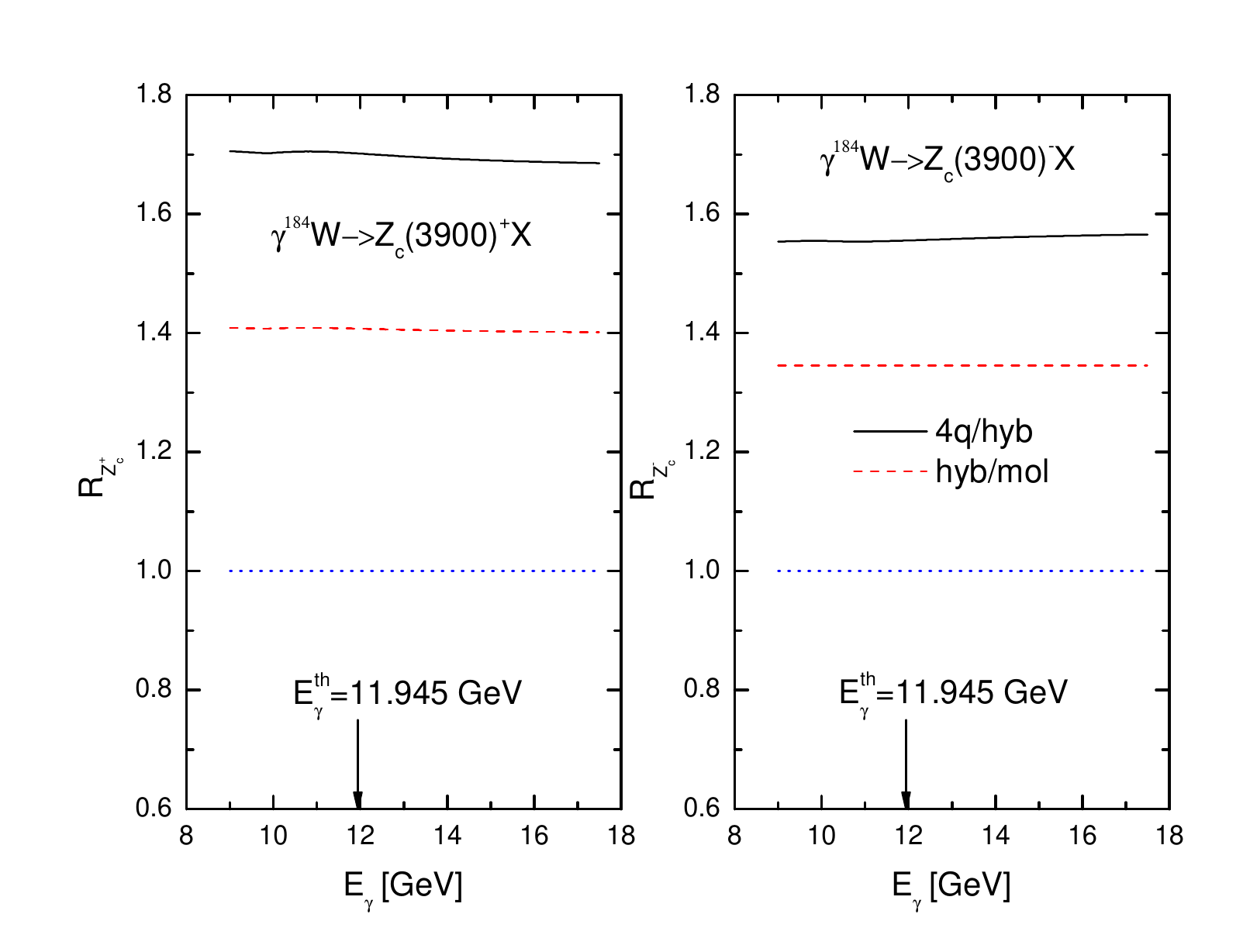}
\vspace*{-2mm} \caption{(Color online.) Ratios between the $Z_c(3900)^{+}$ (left panel) and between the $Z_c(3900)^-$ (right panel) production cross sections on $^{184}$W, shown in Fig. 2 and calculated in the compact tetraquark and hybrid scenarios, in the hybrid and molecular scenarios of the $Z_c(3900)^{\pm}$ for an off-shell target nucleons, as functions of photon energy. The arrows indicate the threshold energy for the $Z_c(3900)^+$ and $Z_c(3900)^-$ photoproduction on a free nucleons in reactions (1) and (5), respectively.}
\label{void}
\end{center}
\end{figure}
\begin{figure}[!h]
\begin{center}
\includegraphics[width=15.0cm]{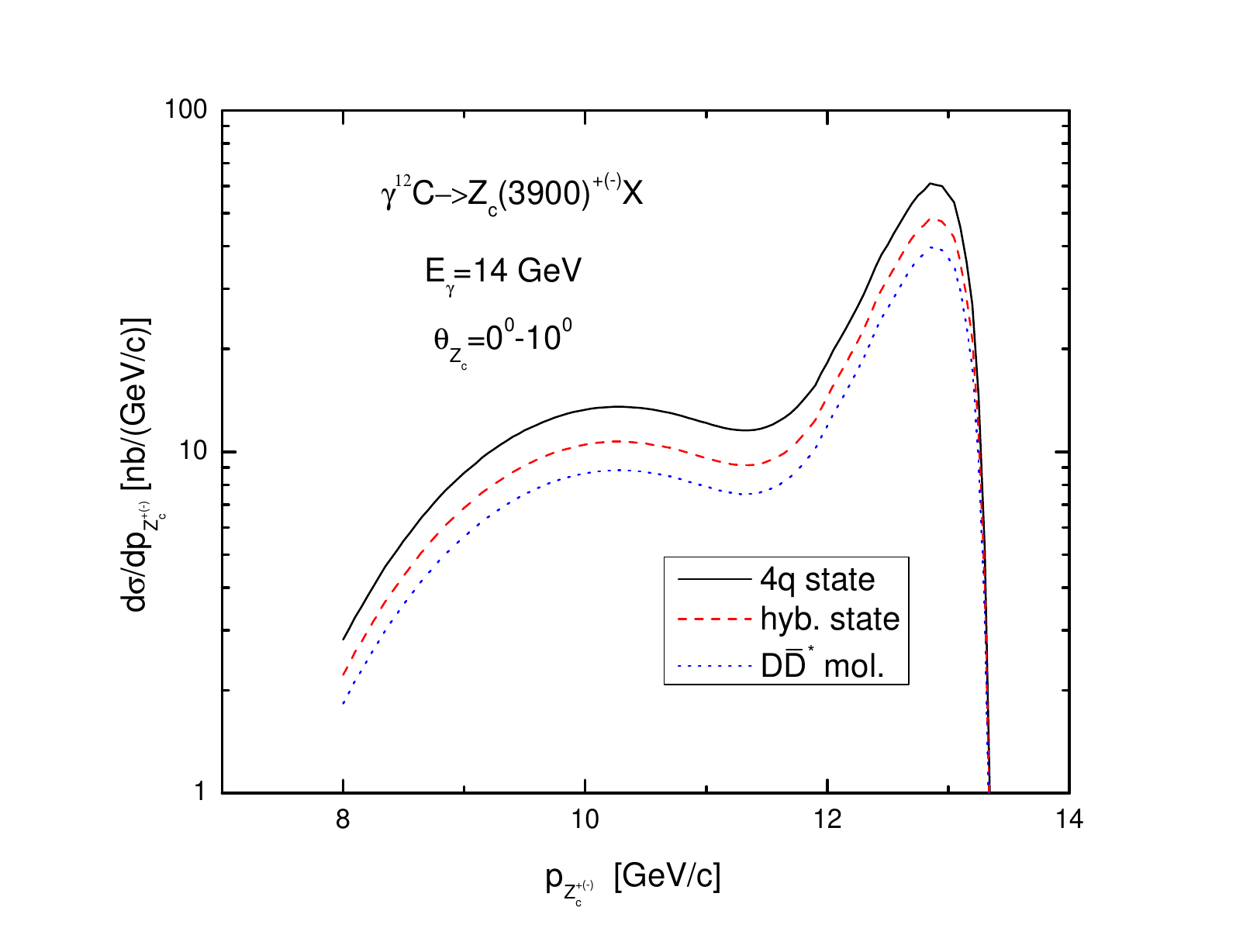}
\vspace*{-2mm} \caption{(Color online.) Momentum differential cross sections for the production of
$Z_c(3900)^+$ and $Z_c(3900)^-$ mesons, respectively, from the direct reactions
(1)--(4) and (5)--(8) proceeding on an off-shell target nucleons in the laboratory polar angular range of 0$^{\circ}$--10$^{\circ}$ in the interaction of photons with energy of $E_{\gamma}=$ 14 GeV with $^{12}$C target nucleus in the considered theoretical pictures describing their intrinsic structures.}
\label{void}
\end{center}
\end{figure}
\begin{figure}[!h]
\begin{center}
\includegraphics[width=15.0cm]{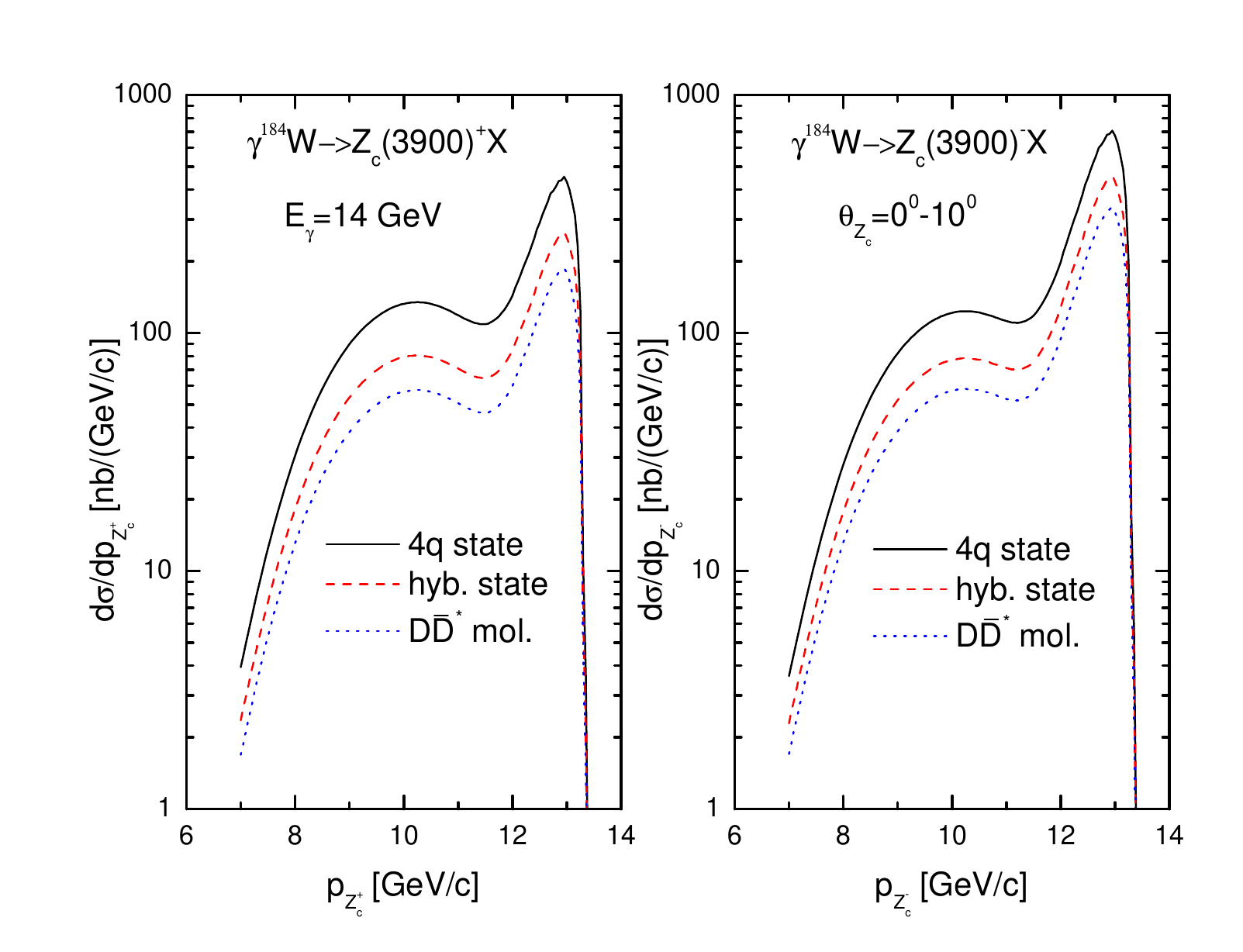}
\vspace*{-2mm} \caption{(Color online.) Momentum differential cross sections for the production of
$Z_c(3900)^+$ (left panel) and $Z_c(3900)^-$ (right panel) mesons, respectively, from the direct reactions
(1)--(4) and (5)--(8) proceeding on an off-shell target nucleons in the laboratory polar angular range of 0$^{\circ}$--10$^{\circ}$ in the interaction of photons with energy of $E_{\gamma}=$ 14 GeV with $^{184}$W target nucleus in the considered theoretical pictures describing their intrinsic structures.}
\label{void}
\end{center}
\end{figure}
\begin{figure}[!h]
\begin{center}
\includegraphics[width=15.0cm]{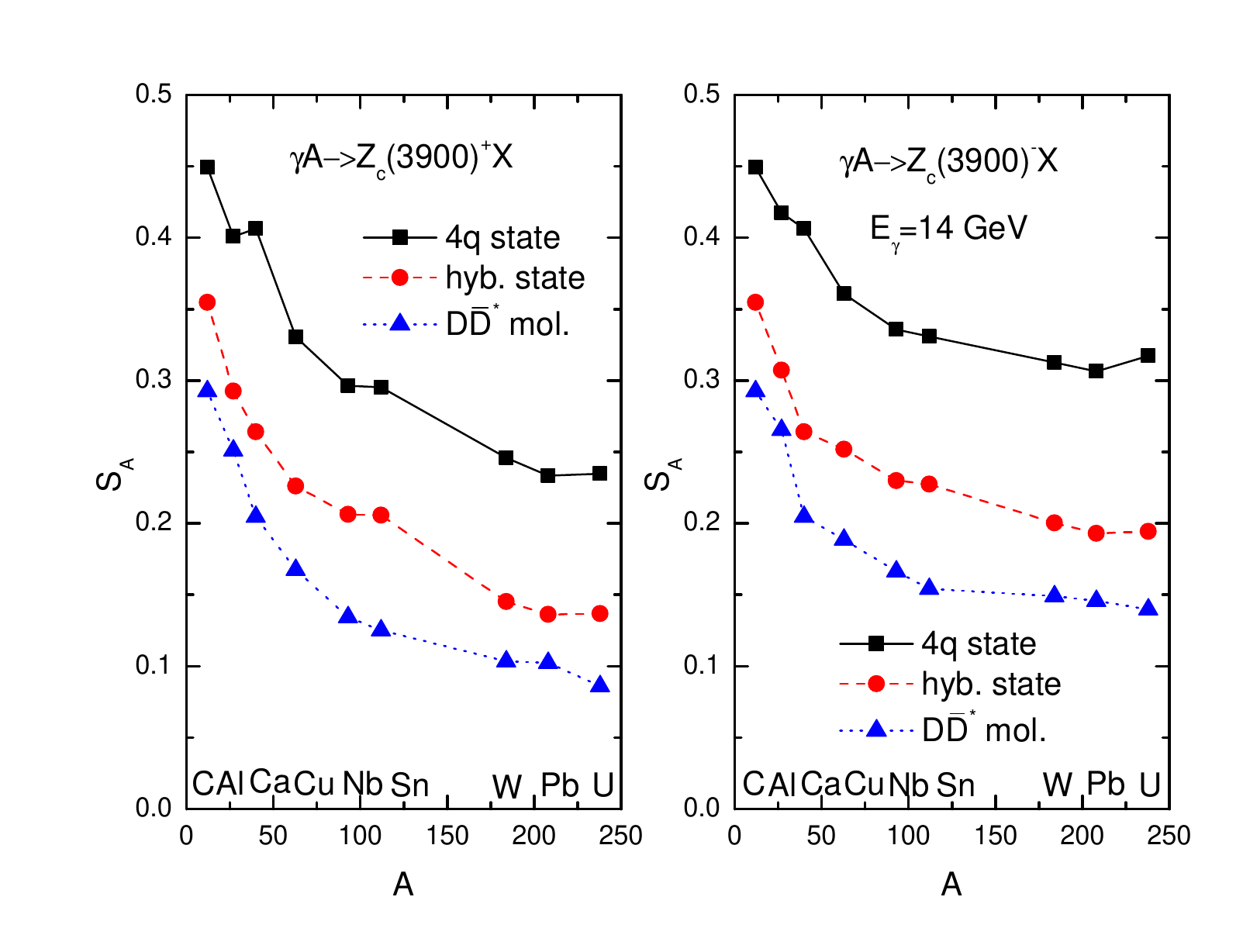}
\vspace*{-2mm} \caption{(Color online.) Transparency ratios $S_A$ for the $Z_c(3900)^+$ (left panel) and $Z_c(3900)^-$ (right panel) mesons, respectively, from the direct processes (1)--(4) and (5)--(8) proceeding on an off-shell target nucleons at incident photon energy of 14 GeV in the laboratory system as functions of the nuclear mass number $A$ in the
considered theoretical pictures describing their intrinsic structures. The lines are to guide the eyes.}
\label{void}
\end{center}
\end{figure}
\begin{figure}[!h]
\begin{center}
\includegraphics[width=15.0cm]{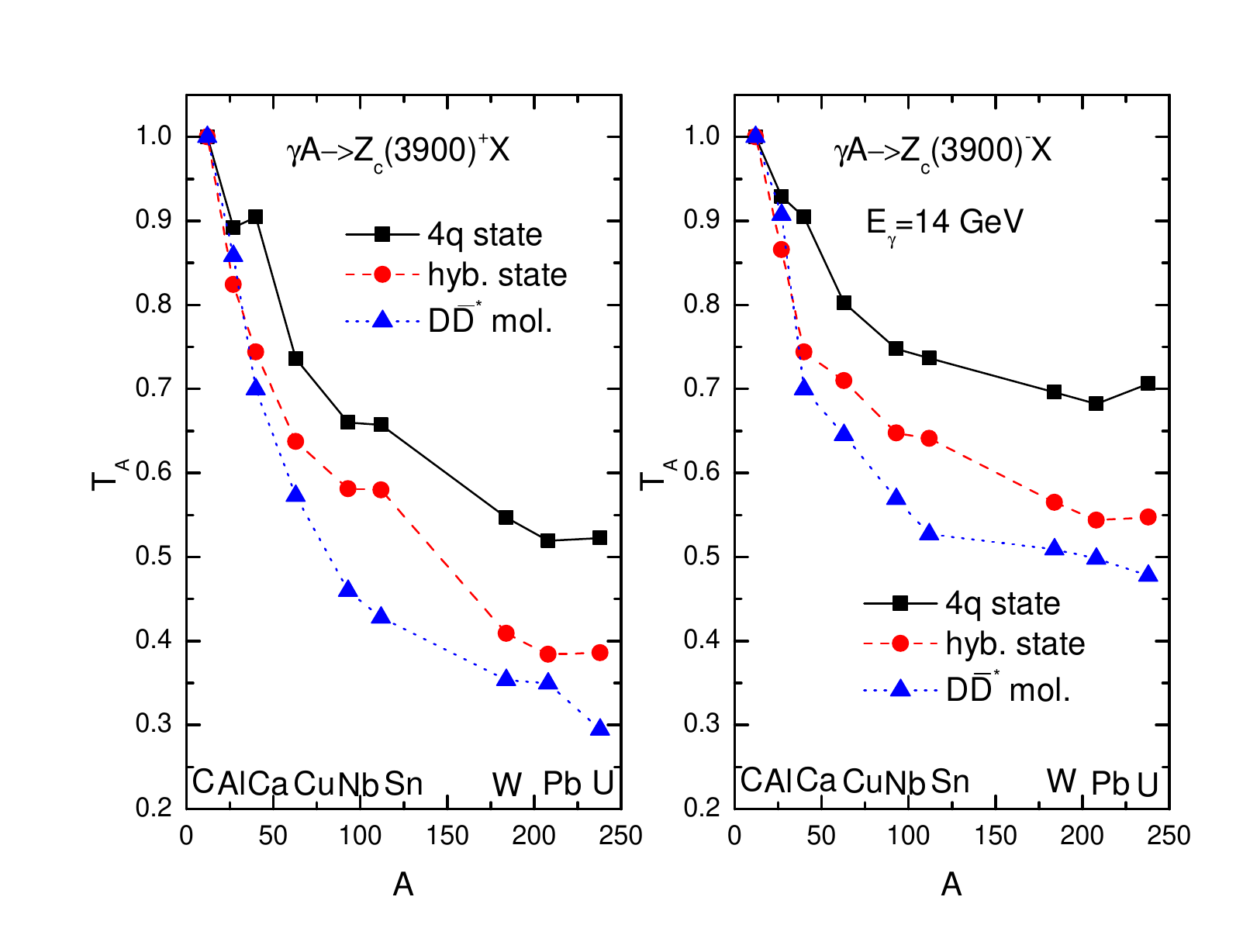}
\vspace*{-2mm} \caption{(Color online.) The same as in Fig. 7, but for the transparency ratios $T_A$.}
\label{void}
\end{center}
\end{figure}
\begin{figure}[!h]
\begin{center}
\includegraphics[width=15.0cm]{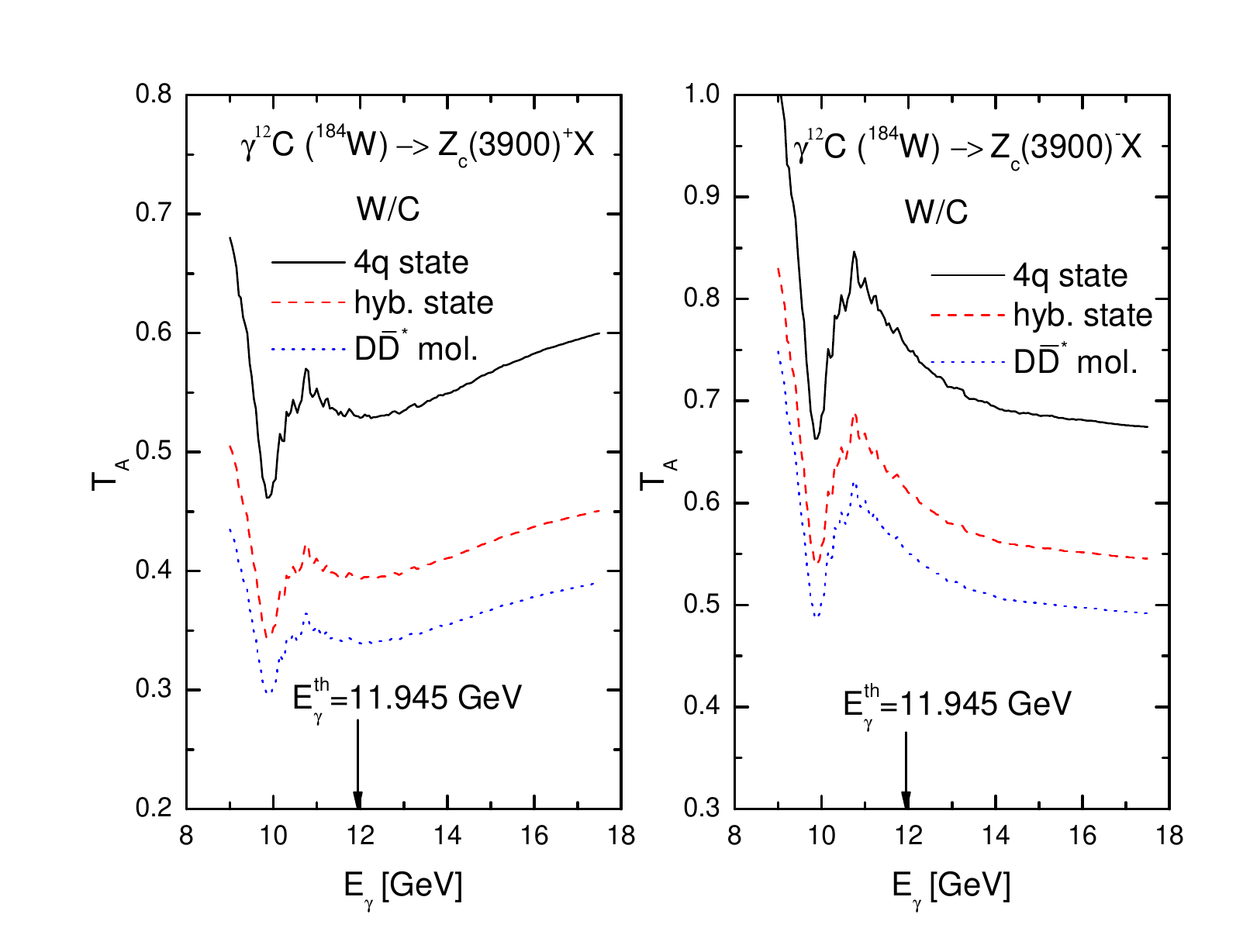}
\vspace*{-2mm} \caption{(Color online.) Transparency ratios $T_A$ for the $Z_c(3900)^+$ (left panel) and
$Z_c(3900)^-$ (right panel) mesons, respectively, from the direct processes (1)--(4) and (5)--(8) proceeding on an off-shell target nucleons as functions of the incident photon energy for combination $^{184}$W/$^{12}$C in the
considered theoretical pictures describing their intrinsic structures. The arrows indicate the threshold energy for the $Z_c(3900)^+$ and $Z_c(3900)^-$ photoproduction on a free nucleons in reactions (1) and (5), correspondingly.}
\label{void}
\end{center}
\end{figure}
\begin{figure}[!h]
\begin{center}
\includegraphics[width=15.0cm]{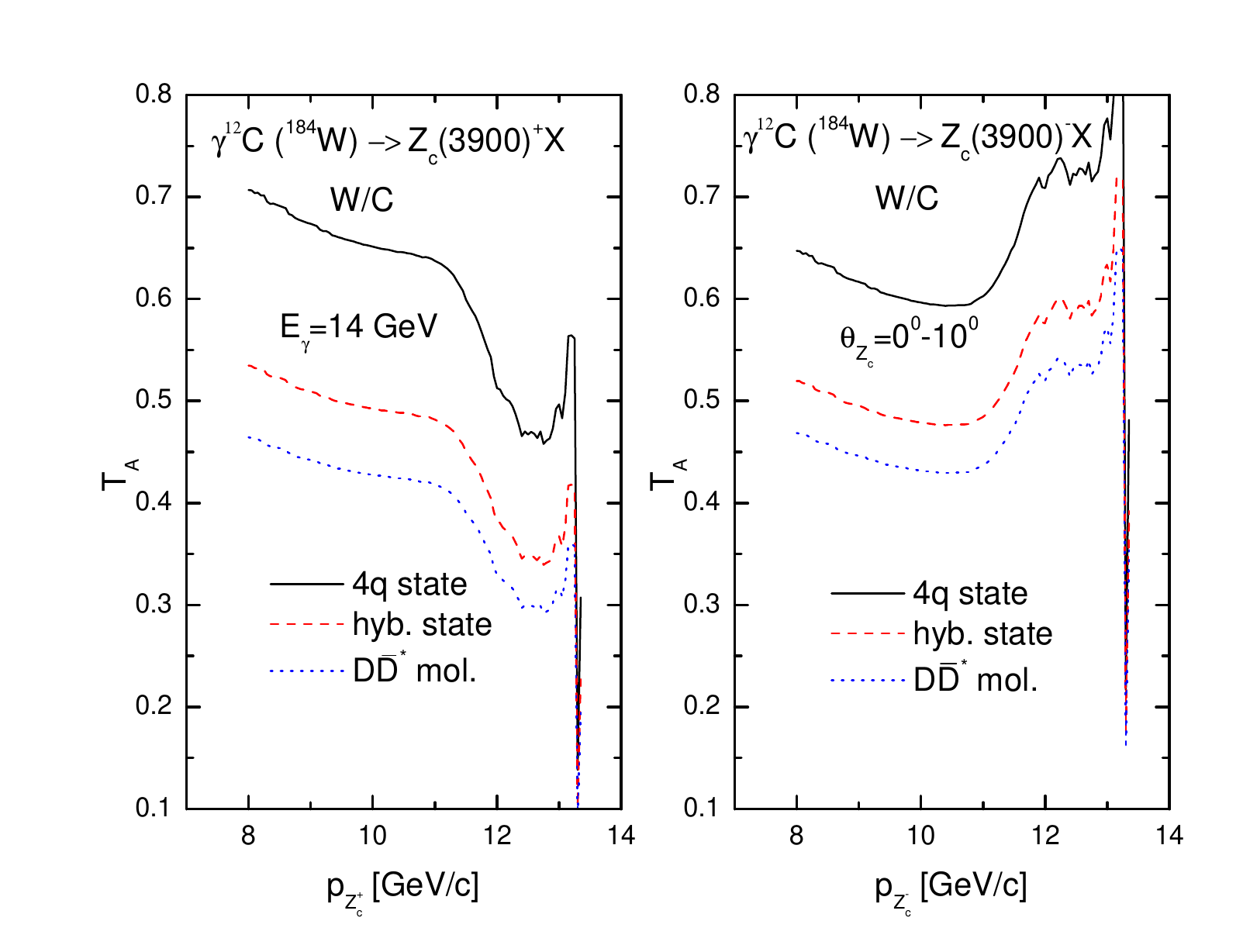}
\vspace*{-2mm} \caption{(Color online.) Transparency ratios $T_A$ for the $Z_c(3900)^+$ (left panel) and
$Z_c(3900)^-$ (right panel) mesons, respectively, from the direct processes (1)--(4) and (5)--(8) proceeding on an off-shell target nucleons as functions of their laboratory momenta for incident photon energy of 14 GeV for combination $^{184}$W/$^{12}$C, calculated in the laboratory polar angular range of 0$^{\circ}$--10$^{\circ}$
in the considered theoretical pictures describing their intrinsic structures.}
\label{void}
\end{center}
\end{figure}

\section*{3. Predictions}

\hspace{1.5cm} The excitation functions for production of $Z_c(3900)^{\pm}$ mesons on $^{12}$C and $^{184}$W target nuclei, calculated in line with Eqs. (9), (10) for three adopted pictures for their intrinsic structures (compact tetraquarks, $D{\bar D}^*$ hadronic molecules and the superpositions of both of them in which there are 50\% of the 4$q$ components and 50\% molecular components) as well as for an off-shell intranuclear target nucleons and for a free ones being at rest, are shown in Figs. 1 and 2, respectively.
One can see that the difference between calculations with and without accounting for the medium effects
(the target nucleon binding and Fermi motion) is sufficiently small at well above threshold photon beam energies
$\sim$ 13--17 GeV, while at lower beam energies their impact on the $Z_c(3900)^{\pm}$ yields is substantial.
It is also nicely seen yet that for both target nuclei considered and for all photon energies studied these yields
show a definite sensitivity to the $Z_c(3900)^{\pm}$ inner configurations.
Thus, we observe for $^{184}$W a well distinguishable and experimentally measurable
changes between the considered scenarios for these configurations: between pure molecular $D{\bar D}^*$ states
and hybrid states with the non-molecular and molecular probabilities of 50\% and 50\%, between
hybrid states with the non-molecular and molecular probabilities of 50\% and 50\% and pure
compact tetraquarks states. They are $\sim$ 40 and 70\% for the $Z_c(3900)^+$ and $\sim$ 35 and 55\% for the $Z_c(3900)^-$. For the light target nucleus $^{12}$C, the sensitivity of the $Z_c(3900)^{\pm}$ production
cross sections to their internal structures becomes lower and, respectively, the same changes as above become somewhat smaller. In line with the aforementioned, they are $\sim$ 21 and 27\% both for the $Z_c(3900)^+$ and for the $Z_c(3900)^-$ and will also be experimentally accessible as well in the future high-precision experiments at the CEBAF facility. More clear information about the sensitivity of the cross sections to the $Z_c(3900)^{\pm}$ intrinsic configurations is contained in Figs. 3 and 4, where the initial photon energy dependences of the ratios $R_{Z_c^{\pm}}$ between the cross sections presented, respectively, in Figs. 1 and 2 are given. In the case of the $Z_c(3900)^+$ production, the ratios $R_{Z_c^{+}}$ being considered take the values of $R_{Z_c^{+}}(4q/{\rm hyb})=1.266$ and 1.706,
$R_{Z_c^{+}}({\rm hyb}/{\rm mol})=1.213$ and 1.409 for carbon and tungsten target nuclei, respectively. The analogous ratios $R_{Z_c^{-}}$ for the $Z_c(3900)^-$ are 1.266 and 1.554, 1.213 and 1.345.
At the same time, the absolute values of the total cross sections have at above threshold photon energies of $\sim$ 13--17 GeV a well measurable strengths $\sim$ 50--200 nb and 200--1500 nb for $Z_c(3900)^+$ and $\sim$ 50--200 nb and 300--2000 nb for $Z_c(3900)^-$ for carbon and tungsten target nuclei, correspondingly.
To stimulate such measurements at the JLab, it is desirable to estimate the $Z_c(3900)^{\pm}$ production rates (the event numbers) in the ${\gamma}^{12}$C($^{184}$W) reactions. For this aim, we translate the $Z_c(3900)^{\pm}$ photoproduction total cross sections, reported above, into the expected event numbers for the $Z_c(3900)^{\pm}$ from the reactions ${\gamma}^{12}{\rm C}(^{184}{\rm W}) \to Z_c(3900)^{\pm}X$, $Z_c(3900)^{\pm} \to {J/\psi}\pi^{\pm}$, $J/\psi \to l^+l^-$
\footnote{$^)$The lepton pairs $l^+l^-$ denote both $\mu^+\mu^-$ and $e^+e^-$.}$^)$.
To estimate the total numbers of the $Z_c(3900)^{\pm}$ events in a one-year run at the CEBAF facility, one needs
to multiply the given above $Z_c(3900)^{\pm}$ photoproduction total cross sections
on the carbon and tungsten target nuclei by the integrated luminosity of $\sim$ 500 pb$^{-1}$ [67]
as well as by the experimental detection efficiency and by the appropriate branching ratios $Br[Z_c(3900)^{\pm} \to {J/\psi}\pi^{\pm}]\approx$ 50\%
\footnote{$^)$The obtained in Refs. [43, 73, 79, 81, 94] results for the $Z_c(3900)^+$ total width and for its partial decay width to ${J/\psi}\pi^+$ indicate that the branching fraction of its decay to the ${J/\psi}\pi^+$ mode is of
about 45-65\%. Relying on this, we will adopt in our exploratory study the conservative (and realistic) value of 50\%
for it. We assume also here that $Br[Z_c(3900)^{-} \to {J/\psi}\pi^{-}]=Br[Z_c(3900)^{+} \to {J/\psi}\pi^{+}]\approx$ 50\%.}$^)$
and $Br[J/\psi \to l^+l^-]\approx$ 12\%. Using a conservative detection efficiency of 10\% for both charged states, we estimate 1.5$\cdot$10$^5$--6.0$\cdot$10$^5$ and 6.0$\cdot$10$^5$--4.5$\cdot$10$^6$ event numbers per year for the $Z_c(3900)^{+}$ signals in the cases of the $^{12}$C and $^{184}$W target nuclei, respectively. The analogous event
numbers for the $Z_c(3900)^{-}$ signals are 1.5$\cdot$10$^5$--6.0$\cdot$10$^5$ and 9.0$\cdot$10$^5$--6.0$\cdot$10$^6$. We see that a sufficiently large number of $Z_c(3900)^{\pm}$  events
\footnote{$^)$Especially $Z_c(3900)^{-}$  events and for the heavy $^{184}$W target nucleus.}$^)$
could be observed at above threshold photon energies of 13--17 GeV for both nuclear targets considered. Therefore, the excitation function measurements for the exotic $Z_c(3900)^{\pm}$ hadrons both on light and heavy target nuclei at upgraded electron accelerator CEBAF-22 will provide a good platform to deepen our understanding of their internal structures
\footnote{$^)$It is worthy pointing out that the behaviors of $Z_c(3900)$ at finite temperature in both the
$D{\bar D}^*$ hadronic molecular and the triangle singularity pictures have been studied in recent publication [118]. It was concluded here that, not only in vacuum but also in hot medium, the behaviors of $Z_c(3900)$ are similar in these two different interpretations of $Z_c(3900)$, i.e., the lineshape of the $Z_c(3900)$ at various temperatures in both these interpretations is almost the same. This implies that it cannot be used in non-vacuum environment for their distinguishing.}$^)$.

The absolute momentum distributions of the $Z_c(3900)^{+}$ and $Z_c(3900)^{-}$ mesons from direct processes
(1)--(8) in the ${\gamma}^{12}$C and ${\gamma}^{184}$W reactions, calculated on the basis of Eqs. (36) and (37) for three adopted scenarios for their inner configurations for laboratory polar angles of 0$^{\circ}$--10$^{\circ}$
and for initial photon energy of 14 GeV, are shown, respectively, in Figs. 5 and 6. The absolute values of the
momentum distributions of the $Z_c(3900)^{+}$ mesons have a well measurable strengths $\sim$ 10--50 nb/(GeV/c) and 50--500 nb/(GeV/c) in the central momentum region of 9--13 GeV/c in the cases of the $^{12}$C and $^{184}$W target nuclei, respectively. The analogous values for the $Z_c(3900)^{-}$ hadrons in this momentum region are about of 10--50 nb/(GeV/c) and 50--700 nb/(GeV/c). The differential cross sections considered show a rather sizeable variations,
especially for the heavy target nucleus $^{184}$W
\footnote{$^)$Which are similar to those shown in Figs. 1 and 2 and which are sufficient to distinguish three
adopted scenarios for the $Z_c(3900)^{\pm}$ internal structures.}$^)$
, when going from the hadronic molecular to compact tetraquark treatments of the $Z_c(3900)^{\pm}$ states.
This behavior of the differential cross sections can also be used to decipher the nature of the exotic charged
charmonium-like $Z_c(3900)^{\pm}$ states from comparison the present model calculations with
data expected from experiments at the upgraded up to 22 GeV CEBAF facility.

Figs. 7 and 8 show the A-dependences of the transparency ratios $S_A$ and $T_A$ of $Z_c(3900)^{\pm}$ production from
the direct processes (1)--(8) in photon-induced reactions on nuclei
$^{12}$C, $^{27}$Al, $^{40}$Ca, $^{63}$Cu, $^{93}$Nb, $^{112}$Sn, $^{184}$W, $^{208}$Pb, and $^{238}$U.
They have been calculated for the photon energy of 14 GeV in line with Eqs. (32) and (35), respectively,
and for three adopted interpretations of the $Z_c(3900)^{\pm}$. We see that the transparency ratios $S_A$ and $T_A$
show strong variations as "functions" of these interpretations and of the mass number.
They drop strongly with increasing nuclear mass number $A$ and the transparency ratios $S_A$ and $T_A$ reach values
of the order of 0.1 and 0.4, correspondingly, for heavy nuclei like $^{208}$Pb and $^{238}$U in the $D{\bar D}^*$ hadronic molecular scenario -- a large deviation from unity which should be easily seen in a future dedicated experiment. The sensitivity of the transparency ratios $S_A$ and $T_A$ to the $Z_c(3900)^{\pm}$ intrinsic structures is nicely seen in Figs. 7 and 8. Thus, there are a sizeable and measurable changes $\sim$ 23, 45\% in the ratio $S_A$ both for the
$Z_c(3900)^{+}$ and for the $Z_c(3900)^{-}$ between calculations performed assuming for $Z_c(3900)^{\pm}$ pure molecular and hybrid scenarios, hybrid and tetraquark pictures for relatively "light" nuclei ($^{27}$Al,$^{40}$Ca). For the medium-mass ($^{93}$Nb,$^{112}$Sn) and heavy ($^{184}$W,$^{238}$U) target nuclei these changes are in general even larger. They are about 50, 45\% and 45, 65\%, respectively. For the quantity $T_A$ the analogous changes are smaller but yet are experimentally distinguishable in the range of medium and large A. They are about 5, 15\%, 25, 15\% and 18, 30\%, correspondingly, in the cases of "light", medium-mass and heavy target nuclei mentioned above.
Therefore, one can conclude that the observation of the A-dependences of the transparency ratios $S_A$ and $T_A$, at least, for medium and large mass numbers $A$ in the future experiments at JLab upgraded to 22 GeV would allow to discriminate between compact tetraquarks, hybrid and molecular interpretations of exotic QCD $Z_c(3900)^{\pm}$ states
and thus to elucidate their nature.

Additional sources of information about this nature are shown in Figs. 9 and 10 the photon energy and meson momentum dependences of the transparency ratio $T_A$ for $Z_c(3900)^{\pm}$ mesons for the $^{184}$W/$^{12}$C combination calculated in line with Eq. (35), using the results presented in Figs. 1, 2 and 5, 6, respectively
\footnote{$^)$Calculating the "differential" transparency ratios $T_A$ for $Z_c(3900)^{\pm}$ mesons, presented in Fig. 10, we replace in Eq. (35) the total (integral) $Z_c(3900)^{\pm}$ production cross sections (9), (10) by the differential ones (36), (37).}$^)$.
One can see that the sensitivity of both dependences to different interpretations of $Z_c(3900)^{\pm}$ mesons is similar to that available in Fig. 8 for heavy nuclei. Therefore, they can also be used for discriminating between possible scenarios for their internal structures.

In the end, we evaluate the theoretical uncertainties of our absolute and relative calculations, reported above, associated mainly with the experimentally unknown total cross section for ${\gamma}p \to Z_c^+n$ reaction giving
the main contribution to the $Z_c^+$ yields from nuclei at near-threshold photon energies. For this aim, we adopted
for this cross section also that corresponding to the total cross section of the ${\gamma}p \to {J/\psi}{\pi^+}n$
process presented in Fig. 8 of Ref. [58]  for the realistic partial decay width $\Gamma[Z_c^+ \to {J/\psi}{\pi^+}]=29$
MeV and corrected for the respective branching ratio $Br[Z_c^+ \to {J/\psi}{\pi^+}]=$ 63\%. By comparing this cross
section  with that from Refs. [59, 60] (Eq. (14)), we find that their behaviors are quantitatively similar to each
other at c.m. energies $W_{\rm th} \le W \le 5.0$ GeV, while for energies $5.0 < W \le 5.8$ GeV the former cross
section is larger than the latter one by a factor about 1.5--2.7. The replacement of the elementary cross section
(14) by that from Ref. [58] leads, as we found, to an enhancement of the absolute $Z_c^+$ production cross sections
on the considered target nuclei by factor of about 1.3--1.5 at photon laboratory energies 13--17 GeV and to a reduction
of the transparency ratios $S_A$ and $T_A$ by only the small fractions of about 10--15\% and 1--3\%, respectively.
In view of the above-mentioned, this means that we can definitely use the parametrization (14) for the total cross
section of the ${\gamma}p \to Z_c^+n$ reaction to provide certain guidance for the measurements of the absolute
yields of $Z_c^+$ mesons from nuclei in future photoproduction experiments and to impose some (preliminary) constraints on their internal structure. To put a strong constraints on this structure using the $Z_c^+$ absolute yields one needs
to know precisely the proton-target $Z_c^+$ production cross sections. On the other hand, since the calculated
differences between the considered scenarios for the $Z_c^+$ inner configuration for the quantities $S_A$ and $T_A$,
in line with the aforementioned, are substantially bigger than the above fractions, the use of the obtained in the
present work predictions for them for discriminating between these scenarios is sufficiently well justified. The
analogous conclusions can be also made for the $Z_c^-$ state.

Finally, we conclude that the absolute and relative observables (integral and differential) considered in the present work can help elucidate the genuine $Z_c(3900)^{\pm}$ intrinsic structures.

\section*{4. Summary and outlook}

\hspace{1.5cm} In this paper, we have investigated the possibility to study the charged charmonium-like state $Z_c(3900)$ production off nuclear targets and its intrinsic structure in inclusive near-threshold photon-induced reactions within the collision model based on the nuclear spectral function. The model accounts for its charged components $Z_c(3900)^{\pm}$ production in direct photon--nucleon interactions as well as three different popular scenarios for their internal structures: compact tetraquarks, hadronic molecules formed by pairs of charmed and anticharmed mesons,
and mixtures of both of them. We have calculated the absolute and relative excitation functions for production of $Z_c(3900)^{\pm}$ mesons on $^{12}$C and $^{184}$W target nuclei at initial photon energies of 9.0--17.5 GeV, the absolute momentum differential cross sections and their ratios for the $Z_c(3900)^{\pm}$ production off these target nuclei at laboratory polar angles of 0$^{\circ}$--10$^{\circ}$ and at photon energy of 14 GeV as well as the A-dependences of the ratios of the total cross sections for $Z_c(3900)^{\pm}$ production at photon energy of 14 GeV within the adopted scenarios for the $Z_c(3900)^{\pm}$ internal structures. We show that the absolute and relative observables considered reveal a certain sensitivity to the still unknown $Z_c(3900)^{\pm}$ internal structures. Therefore, they might be useful for determination of these structures -- the issue which has attracted considerable interest in recent years -- from the comparison of them with the experimental data from the future high-precision experiments at the upgraded up to 22 GeV CEBAF facility. We hope that the findings of the present work will additionally stimulate the conducting here such experiments.
\\

\end{document}